\documentclass[preprint,final,5p,times,twocolumn]{elsarticle}
\usepackage{rotating,color,subfigure,amssymb}
\usepackage{alphalph}
\usepackage{amsmath}
\usepackage[T1]{fontenc}
\usepackage{amssymb}
\journal{Physics Letters B}
\begin{document}
                                       
\begin{frontmatter}

\title{ Search for an Isospin $I =3$ Dibaryon}
\date{\today}

\author[IKPUU]{The WASA-at-COSY Collaboration\\[2ex] P.~Adlarson\fnref{fnmz}}
\author[ASWarsN]{W.~Augustyniak}
\author[IPJ]{W.~Bardan}
\author[Edinb]{M.~Bashkanov\corref{coau}}\ead{mikhail.bashkanov@ed.ac.uk}
\author[MS]{F.S.~Bergmann}
\author[ASWarsH]{M.~Ber{\l}owski}
\author[IITB]{H.~Bhatt}
\author[Budker,Novosib]{A.~Bondar}
\author[IKPJ]{M.~B\"uscher\fnref{fnpgi,fndus}}
\author[IKPUU]{H.~Cal\'{e}n}
\author[IPJ]{I.~Ciepa{\l}}
\author[PITue,Kepler]{H.~Clement}
\author[IPJ]{E.~Czerwi{\'n}ski}
\author[MS]{K.~Demmich}
\author[IKPJ]{R.~Engels}
\author[ZELJ]{A.~Erven}
\author[ZELJ]{W.~Erven}
\author[Erl]{W.~Eyrich}
\author[IKPJ,ITEP]{P.~Fedorets}
\author[Giess]{K.~F\"ohl}
\author[IKPUU]{K.~Fransson}
\author[IKPJ]{F.~Goldenbaum}
\author[IKPJ,IITI]{A.~Goswami}
\author[IKPJ,HepGat]{K.~Grigoryev\fnref{fnac}}
\author[IKPUU]{C.--O.~Gullstr\"om}
\author[IKPUU]{L.~Heijkenskj\"old}
\author[IKPJ,JCHP]{V.~Hejny}
\author[MS]{N.~H\"usken}
\author[IPJ]{L.~Jarczyk}
\author[IKPUU]{T.~Johansson}
\author[IPJ]{B.~Kamys}
\author[ZELJ]{G.~Kemmerling\fnref{fnjcns}}
\author[IKPJ]{F.A.~Khan}
\author[IPJ]{G.~Khatri}
\author[MS]{A.~Khoukaz}
\author[IPJ]{O.~Khreptak}
\author[HiJINR]{D.A.~Kirillov}
\author[IPJ]{S.~Kistryn}
\author[ZELJ]{H.~Kleines\fnref{fnjcns}}
\author[Katow]{B.~K{\l}os}
\author[IPJ]{W.~Krzemie{\'n}}
\author[IFJ]{P.~Kulessa}
\author[IKPUU,ASWarsH]{A.~Kup{\'s}{\'c}}
\author[Budker,Novosib]{A.~Kuzmin}
\author[NITJ]{K.~Lalwani}
\author[IKPJ]{D.~Lersch}
\author[IKPJ]{B.~Lorentz}
\author[IPJ]{A.~Magiera}
\author[IKPJ,JARA]{R.~Maier}
\author[IKPUU]{P.~Marciniewski}
\author[ASWarsN]{B.~Maria{\'n}ski}
\author[ASWarsN]{H.--P.~Morsch}
\author[IPJ]{P.~Moskal}
\author[IKPJ]{H.~Ohm}
\author[PITue,Kepler]{E.~Perez del Rio\fnref{fnlnf}}
\author[HiJINR]{N.M.~Piskunov}
\author[IKPJ]{D.~Prasuhn}
\author[IKPUU,ASWarsH]{D.~Pszczel}
\author[IFJ]{K.~Pysz}
\author[IKPUU,IPJ]{A.~Pyszniak}
\author[IKPJ,JARA,Bochum]{J.~Ritman}
\author[IITI]{A.~Roy}
\author[IPJ]{Z.~Rudy}
\author[IPJ]{O.~Rundel}
\author[IITB,IKPJ]{S.~Sawant}
\author[IKPJ]{S.~Schadmand}
\author[IPJ]{I.~Sch\"atti--Ozerianska}
\author[IKPJ]{T.~Sefzick}
\author[IKPJ]{V.~Serdyuk}
\author[Budker,Novosib]{B.~Shwartz}
\author[MS]{K.~Sitterberg}
\author[PITue,Kepler,Tomsk]{T.~Skorodko}
\author[IPJ]{M.~Skurzok}
\author[IPJ]{J.~Smyrski}
\author[ITEP]{V.~Sopov}
\author[IKPJ]{R.~Stassen}
\author[ASWarsH]{J.~Stepaniak}
\author[Katow]{E.~Stephan}
\author[IKPJ]{G.~Sterzenbach}
\author[IKPJ]{H.~Stockhorst}
\author[IKPJ,JARA]{H.~Str\"oher}
\author[IFJ]{A.~Szczurek}
\author[ASWarsN]{A.~Trzci{\'n}ski}
\author[IITB]{R.~Varma}
\author[IKPUU]{M.~Wolke}
\author[IPJ]{A.~Wro{\'n}ska}
\author[ZELJ]{P.~W\"ustner}
\author[KEK]{A.~Yamamoto}
\author[ASLodz]{J.~Zabierowski}
\author[IPJ]{M.J.~Zieli{\'n}ski}
\author[Erl]{A.~Zink}
\author[IKPUU]{J.~Z{\l}oma{\'n}czuk}
\author[ASWarsN]{P.~{\.Z}upra{\'n}ski}
\author[IKPJ]{M.~{\.Z}urek}

\address[IKPUU]{Division of Nuclear Physics, Department of Physics and 
 Astronomy, Uppsala University, Box 516, 75120 Uppsala, Sweden}
\address[ASWarsN]{Department of Nuclear Physics, National Centre for Nuclear 
 Research, ul.\ Hoza~69, 00-681, Warsaw, Poland}
\address[IPJ]{Institute of Physics, Jagiellonian University, ul.\ Reymonta~4, 
 30-059 Krak\'{o}w, Poland}
\address[Edinb]{School of Physics and Astronomy, University of Edinburgh, 
 James Clerk Maxwell Building, Peter Guthrie Tait Road, Edinburgh EH9 3FD, 
 Great Britain}
\address[MS]{Institut f\"ur Kernphysik, Westf\"alische Wilhelms--Universit\"at 
 M\"unster, Wilhelm--Klemm--Str.~9, 48149 M\"unster, Germany}
\address[ASWarsH]{High Energy Physics Department, National Centre for Nuclear 
 Research, ul.\ Hoza~69, 00-681, Warsaw, Poland}
\address[IITB]{Department of Physics, Indian Institute of Technology Bombay, 
 Powai, Mumbai--400076, Maharashtra, India}
\address[Budker]{Budker Institute of Nuclear Physics of SB RAS, 11~akademika 
 Lavrentieva prospect, Novosibirsk, 630090, Russia}
\address[Novosib]{Novosibirsk State University, 2~Pirogova Str., Novosibirsk, 
 630090, Russia}
\address[PITue]{Physikalisches Institut, Eberhard--Karls--Universit\"at 
 T\"ubingen, Auf der Morgenstelle~14, 72076 T\"ubingen, Germany}
\address[Kepler]{Kepler Center for Astro and Particle Physics, Eberhard Karls 
 University T\"ubingen, Auf der Morgenstelle~14, 72076 T\"ubingen, Germany}
\address[IKPJ]{Institut f\"ur Kernphysik, Forschungszentrum J\"ulich, 52425 
 J\"ulich, Germany}
\address[ZELJ]{Zentralinstitut f\"ur Engineering, Elektronik und Analytik, 
 Forschungszentrum J\"ulich, 52425 J\"ulich, Germany}
\address[Erl]{Physikalisches Institut, Friedrich--Alexander--Universit\"at 
 Erlangen--N\"urnberg, Erwin--Rommel-Str.~1, 91058 Erlangen, Germany}
\address[ITEP]{Institute for Theoretical and Experimental Physics, State 
 Scientific Center of the Russian Federation, Bolshaya Cheremushkinskaya~25, 
 117218 Moscow, Russia}
\address[Giess]{II.\ Physikalisches Institut, Justus--Liebig--Universit\"at 
 Gie{\ss}en, Heinrich--Buff--Ring~16, 35392 Giessen, Germany}
\address[IITI]{Department of Physics, Indian Institute of Technology Indore, 
 Khandwa Road, Indore--452017, Madhya Pradesh, India}
\address[HepGat]{High Energy Physics Division, Petersburg Nuclear Physics 
 Institute, Orlova Rosha~2, Gatchina, Leningrad district 188300, Russia}
\address[HiJINR]{Veksler and Baldin Laboratory of High Energiy Physics, Joint 
 Institute for Nuclear Physics, Joliot--Curie~6, 141980 Dubna, Moscow region, 
 Russia}
\address[Katow]{August Che{\l}kowski Institute of Physics, University of 
 Silesia, Uniwersytecka~4, 40-007, Katowice, Poland}
\address[IFJ]{The Henryk Niewodnicza{\'n}ski Institute of Nuclear Physics, 
 Polish Academy of Sciences, 152~Radzikowskiego St, 31-342 Krak\'{o}w, Poland}
\address[NITJ]{Department of Physics, Malaviya National Institute of 
 Technology Jaipur, 302017, Rajasthan, India}
\address[JARA]{JARA--FAME, J\"ulich Aachen Research Alliance, Forschungszentrum 
 J\"ulich, 52425 J\"ulich, and RWTH Aachen, 52056 Aachen, Germany}
\address[Bochum]{Institut f\"ur Experimentalphysik I, Ruhr--Universit\"at 
 Bochum, Universit\"atsstr.~150, 44780 Bochum, Germany}
\address[Tomsk]{Department of Physics, Tomsk State University, 36~Lenina 
 Avenue, Tomsk, 634050, Russia}
\address[KEK]{High Energy Accelerator Research Organisation KEK, Tsukuba, 
 Ibaraki 305--0801, Japan}
\address[ASLodz]{Department of Astrophysics, National Centre for Nuclear 
 Research, Box~447,   90--950 {\L}\'{o}d\'{z}, Poland}

\fntext[fnmz]{present address: Institut f\"ur Kernphysik, Johannes 
 Gutenberg--Universit\"at Mainz, Johann--Joachim--Becher Weg~45, 55128 Mainz, 
 Germany}
\fntext[fnpgi]{present address: Peter Gr\"unberg Institut, PGI--6 Elektronische 
 Eigenschaften, Forschungszentrum J\"ulich, 52425 J\"ulich, Germany}
\fntext[fndus]{present address: Institut f\"ur Laser-- und Plasmaphysik, 
 Heinrich--Heine Universit\"at D\"usseldorf, Universit\"atsstr.~1, 40225 
 D\"usseldorf, Germany}
\fntext[fnac]{present address: III.~Physikalisches Institut~B, Physikzentrum, 
 RWTH Aachen, 52056 Aachen, Germany}
\fntext[fnjcns]{present address: J\"ulich Centre for Neutron Science JCNS, 
 Forschungszentrum J\"ulich, 52425 J\"ulich, Germany}
\fntext[fnlnf]{present address: INFN, Laboratori Nazionali di Frascati, Via 
 E.~Fermi, 40, 00044 Frascati (Roma), Italy}



\cortext[coau]{Corresponding author }

\begin{abstract}
Various theoretical calculations based on QCD or hadronic interactions predict
that in addition to the recently observed dibaryon resonance $d^*(2380)$ with
$I(J^P) = 0(3^+)$ there should also exist a dibaryon resonance with mirrored
quantum numbers $I(J^P) = 3(0^+)$. We report here on a search for such a
$NN$-decoupled state in data on the $pp \to pp\pi^+\pi^+\pi^-\pi^-$
reaction. Since no clear-cut evidence has been found, we give upper limits for
the production cross section of such a resonance in the mass range 2280 - 2500
MeV. 

\end{abstract}


\end{frontmatter}



\section{Introduction}
Recently, exclusive and kinematically complete measurements of the reactions
$pn \to d \pi^0\pi^0$ and $pn \to d \pi^+\pi^-$ 
revealed a narrow  resonance-like structure in the total cross section
\cite{mb,MB,isofus} at a mass $m \approx$ 2380~MeV with a width of 
$\Gamma \approx$ 70 MeV and quantum numbers $I(J^P) = 0(3^+)$. Additional
evidence for it had been traced subsequently in the two-pion production
reactions $pn \to pp\pi^0\pi^-$ \cite{pp0-}, $pn \to pn\pi^0\pi^0$
\cite{pn00} and $pn \to pn\pi^+\pi^-$ \cite{pn+-,exa,hades}. Finally, analyzing
power measurements of $np$ elastic scattering established this structure to
represent a true $s$-channel resonance, which produces a pole in the $^3D_3$
partial wave --- denoted since then by $d^*(2380)$
\cite{prl2014,pnfull,RW,RWnew}. 

Such a dibaryon resonance, which asymptotically resembles a deeply bound
$\Delta\Delta$ system, was predicted first -- and astonishingly precise as it
turns out now -- by Dyson and Xuong \cite{dyson} in 1964 based on $SU(6)$
symmetry breaking. Later-on, Goldman {\it et al.} \cite{goldman} called this
state the "inevitable dibaryon" pointing out that due to its particular
quantum numbers and its associated special symmetries, such a state must be
predicted in any theoretical model based on confinement and one-gluon
exchange. Indeed, there are now quite a number of QCD-based model
calculations available, which find $d^*(2380)$ at about the correct mass
\cite{ping,shen,zhang1,dong,zhang2,chen,kamae,mulders}. Also, relativistic
Faddeev-type calculations based on hadronic interactions find this state at
the observed mass \cite{GG1,GG2}. The observed relatively narrow width of about
70 MeV is obviously more difficult to understand theoretically. Until
recently, Gal and Garcilazo came closest with about 100 MeV \cite{GG2}. Dong
{\it et al.} \cite{dong} succeeded in reproducing the experimental width 
by accounting for hidden color effects -- as had been speculated already in
Ref. \cite{BBC}. 

Part of the theoretical calculations, which successfully obtain the
$d^*(2380)$ state, predicts also a dibaryon state with mirrored quantum numbers
$I(J^P) = 3(0^+)$ at a similar mass \cite{dyson,ping,shen,GG2}. In tendency,
this state, which is expected to be again a $\Delta\Delta$ configuration
asymptotically, appears to be somewhat less bound than $d^*(2380)$, but still
below the 
$\Delta\Delta$ threshold of $2m_{\Delta}$. The predicted width varies from
about 90 MeV \cite{GG2} to about 180 MeV \cite{ping}. Only the calculation of
the Nijmegen group \cite{mulders} predicts this state to be far above the
$\Delta\Delta$ threshold.

\section{Experiment}

In order to investigate this issue experimentally, we follow the suggestion of
Dyson and Xuong (who called the state in question the $D_{30}$, where the first
index denotes the isospin and the second one the spin) and consider the
four-pion production in proton-proton collisions, in particular the $pp \to
pp\pi^+\pi^+\pi^-\pi^-$ reaction. Because of its isospin $I = 3$ such a state is
isospin-decoupled from the nucleon-nucleon ($NN$) system. Therefore, in order
to be able to reach such a state by $NN$ collisions, its production in the
collision process needs to be associated by the generation of particles, which
take away two units of isospin. This appears to be accomplished most easily
by the production of two extra pions. So the process we aim at reads as $pp
\to D_{30} \pi^-\pi^- \to \Delta^{++}\Delta^{++} \pi^-\pi^- \to
pp\pi^+\pi^+\pi^-\pi^-$. Due to $I$ = 3 the $\Delta^{++}\Delta^{++}$
configuration is the most 
preferred $\Delta\Delta$ combination, where $D_{30}$ decays into -- see next
section. 

The measurements of this reaction have been carried out
with the WASA detector including a hydrogen pellet target \cite{CB,wasa} at
the cooler synchrotron COSY (Forschungszentrum J\"ulich) using proton beams
with energies of $T_p$~=~2.063~and~2.541~GeV. These correspond to
center-of-mass energies of $\sqrt s$ = 2.72 and 2.88 GeV, respectively. The
latter denotes the highest energy used with WASA at COSY.

The trigger was set to two (and more) charged hits both in the forward and in
the central detector. Since the main goals of these runs, which comprise 3
weeks of beam-time, was not the four-pion
production, but $\omega$ and $\eta'$ production, also a missing mass trigger
was active during the measurements at 2.541 GeV. It required the deposited
energy of each of the two
protons detected in the forward detector to be larger than 150 MeV -- a
condition, which did not affect the four-pion production events.

The four-momenta of the two emitted protons were detected in the Forward
Detector, whereas the four-momenta of the four charged pions were recorded in
the Central Detector, where a magnetic field allowed for charge
identification and momentum determination. That way, the reaction was measured
kinematically complete with  
four overconstraints, which allowed a corresponding kinematic fit of the
events. Finally, a total of 136 (1017) events at $T_p$ = 2.063 (2.541) GeV
passed the $\chi^2$ criterion of the kinematic fit, which is a cut of the
probability function at 10$\%$. Despite the kinematic fit
condition, Monte Carlo (MC) simulations suggest that the final sample at $T_p$ =
2.063 GeV is contaminated with about 50 events originating from three-pion
production, where the photons from $\pi^0$ decay have undergone conversion or
Dalitz decay. 
The distribution of those events is not noticeably different from the other
events. 
The acceptance of the WASA detector for $pp\pi^+\pi^+\pi^-\pi^-$ events has
been determined by MC simulations to be about 15$\%$ with an uncertainty of
less than 1$\%$. 
The detection efficiency for $pp\pi^+\pi^+\pi^-\pi^-$ events has been 0.1 $\%$
at  $T_p$ = 2.063 GeV and 0.5$\%$ at  $T_p$ = 2.541 GeV, also evaluated via
comprehensive MC simulations of the detector performance.

The absolute normalization of the four-pion production data has been done via
the simultaneous measurement of the three-pion production ($\pi^+\pi^-\pi^0$
with $\pi^0$ decay into two photons) including $\eta$ and
$\omega$ production in this channel. The spectra of the
three-pion invariant mass $M_{\pi^+\pi^0\pi^-}$ at $T_p$ = 2.063 GeV and $T_p$
= 2.541 GeV are shown in Fig. 1. 
   Since the cross section for three-pion production is two orders of magnitude
   larger than that for four-pion production, it was sufficient to use only a
   small sample of the available three-pion production data for this procedure.

At $T_p$ = 2.063 GeV, the data for the three-pion production (including $\eta$
and $\omega$ production) have been normalized to the value of 220 $\mu$b
obtained in Refs. \cite{Pickup1,Pickup2} for this reaction at $T_p$ = 2 GeV. As
a result, the fitted contributions from $\eta$ and $\omega$ production (dotted
and dashed lines in Fig. 1) correspond to production cross sections of
$111 \pm 20~\mu$b and $5.6 \pm 1.0~\mu$b, respectively. The
first value agrees reasonably well with the value of $142 \pm 22~\mu$b obtained at
$T_p$ = 2.2 GeV by the HADES collaboration \cite{Hades}. The second value
is in good agreement with the value of 5.7 $\mu$b obtained in
Ref. \cite{Barsov}. 

At $T_p$ = 2.541 GeV, where $\omega$ production provides already a substantial
contribution to three-pion production, the data have been
normalized to this process using for the $\omega$ production cross section the
value 35 $\mu$b interpolated from the values given in Refs. \cite{TOF,DISTO}. 

As a result of this normalization procedure, we obtain total four-pion cross
sections of $0.7 \pm 0.3$ and $5.2 \pm 1.0~\mu$b at $T_p$~=~2.063 and 2.541 GeV,
respectively, which are two orders of magnitude smaller than the 
three-pion production cross sections at these energies.

The quoted uncertainties originate predominantly from systematic uncertainties
in the determination of background beneath the $\eta$ and $\omega$ peaks, the
determination of the WASA acceptance and efficiency and the extrapolation of
cross sections to full phase space, which has been done by MC simulations
assuming phase-space or model distributions -- see next section.

   Due to the small statistics for the four-pion production reaction, all
   systematical effects had to be evaluated by Monte Carlo simulations
   only. However, some of these systematic effects, like influence  
   of missing-mass trigger could be cross-checked with the $pp \to
   pp\pi^+\pi^0\pi^-$ reaction due to much higher statistics, same
   multiplicity  ($\pi^0$ decays into two photons, hence there are also six
   particles detected in the final  state) and very similar kinematics. 
   Therefore systematical errors related to triggering, errors parameterization,
   kinematical fitting, etc. were evaluated based on large samples of three
   pion production data.

\begin{figure} [t]
\centering
\includegraphics[width=0.99\columnwidth]{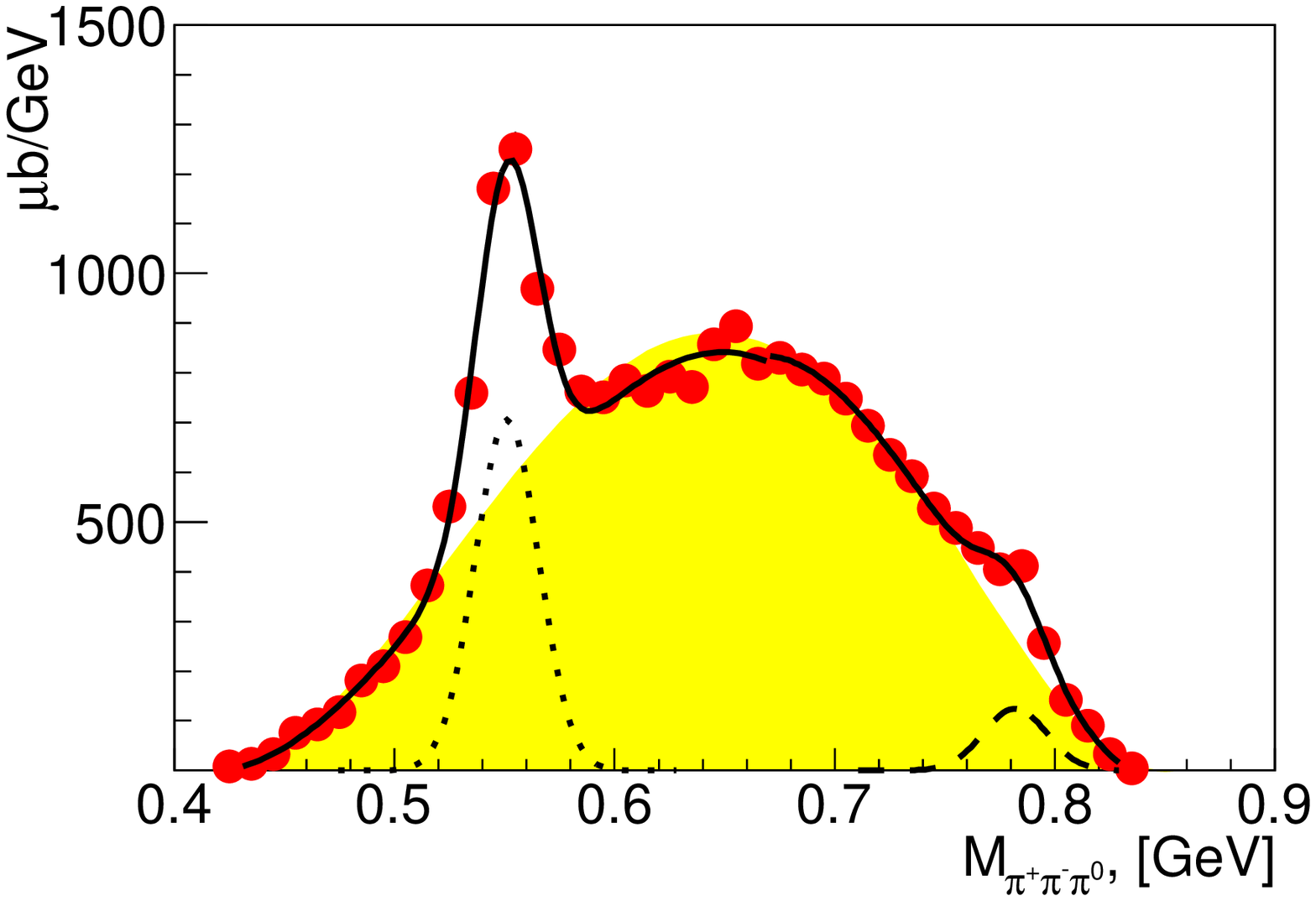}
\includegraphics[width=0.99\columnwidth]{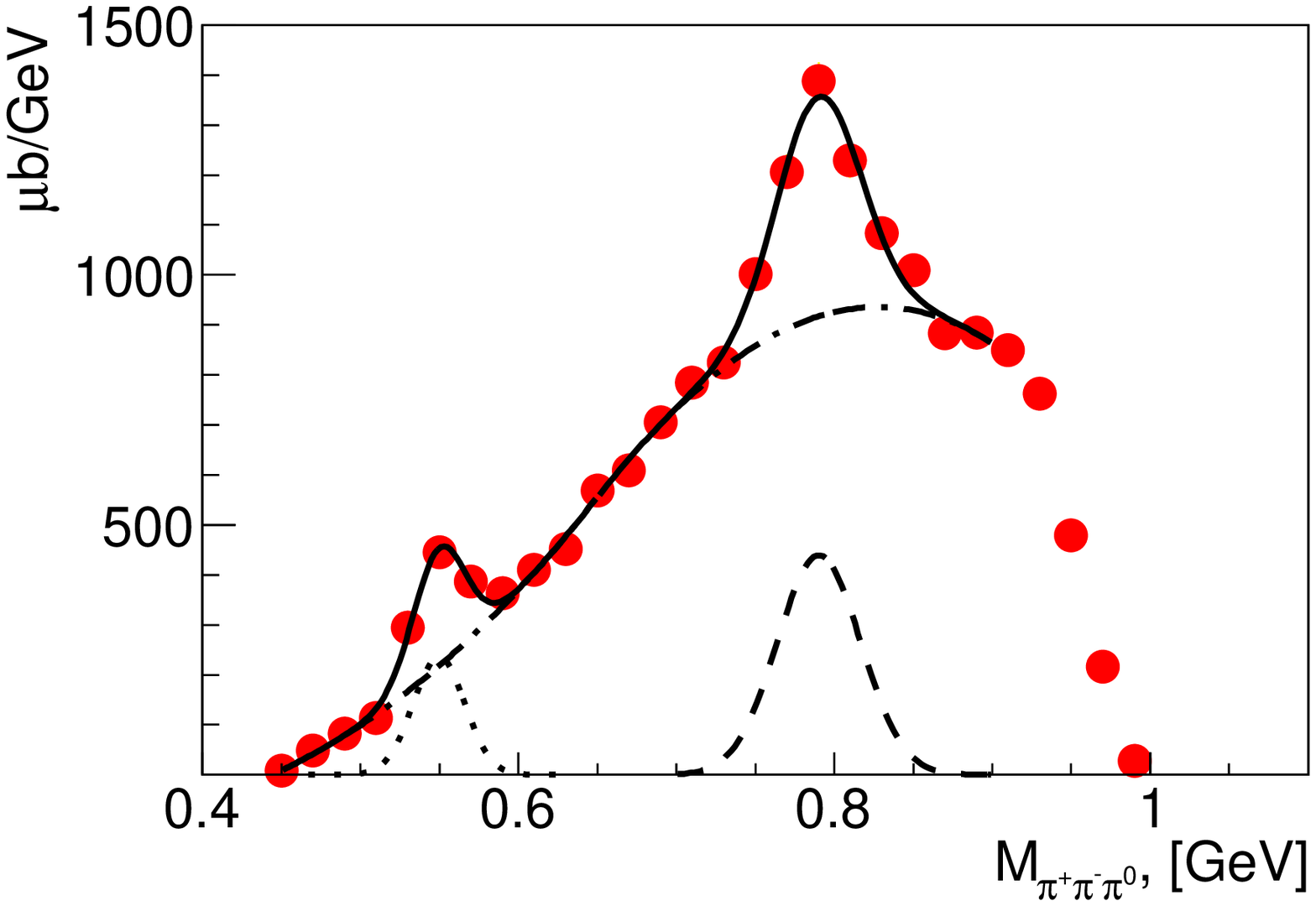}
\caption{\small (Color online) Measured spectrum of the $pp \to
  pp\pi^+\pi^0\pi^-$ reaction at $T_p$ = 2.063 (top) and 2.541 GeV (bottom).  The filled circles represent
  the data from this  
  work. Dashed and dotted lines give the fitted $\eta$ and $\omega$
  contributions, whereas the shaded area in the top panel shows the pure
  phase-space distribution. The dash-dotted line in the bottom panel is a
  polynomial fit of direct three-pion production. The solid line is the sum of
  $\eta$, $\omega$ 
  and direct three-pion production.}
\label{fig1}
\end{figure}

The widths of about 30 MeV for the $\eta$ and $\omega$ lines in the three-pion
spectrum give a measure of the mass resolution achieved in this data
analysis.

\section{Results and Discussion}

The energy dependence of the total cross section for the $pp \to
pp\pi^+\pi^+\pi^-\pi^-$ reaction is displayed in Fig. 2, where our results are
compared to earlier published data \cite {alexan,danieli,ward,blob,blobel}
obtained a 
higher energies. The solid line represents the energy dependence of pure phase
space. It accounts at least qualitatively for the trend of the data.

Having measured the four-pion production kinematically complete, we are able
to 
construct all kinds of differential distributions. Of relevance in the search
for the $I = 3$ dibaryon $D_{30}$ are the spectra of the $pp\pi\pi$-invariant
masses $M_{pp\pi^+\pi^+}$, $M_{pp\pi^-\pi^-}$ and $M_{pp\pi^+\pi^-}$. For the
latter the statistics quadruples due to combinatorics. These spectra are
displayed in Fig.~3 for both incident energies. They are shown within WASA
acceptance, {\it i.e.} not acceptance corrected, in order to avoid any model
dependence introduced by the corresponding correction procedure. 

The spectra in Fig.~3 show very smooth mass
distributions and do not exhibit any unusual structures. However, they deviate
systematically from pure six-body phase-space distributions, which are shown
( again within WASA acceptance) by the shaded histograms in Fig.~3. This is
not surprising, since already
single- and two-pion productions are known to be dominated by baryon excitations
starting right from threshold \cite{Teis,JZ,ED,Luis,IHEP,TS,isopppipi}. 

The lowest-lying baryon resonance,
which decays by emission of two pions, is the Roper resonance
$N^*(1440)$ with its two-pion decay routes $N^* \to N\sigma \to N\pi\pi$ and
$N^* \to \Delta\pi \to N\pi\pi$. It is known to dominate the two-pion
production for energies $T_p 
<$~1~GeV -- before the $\Delta\Delta$ excitation by $t$-channel meson exchange
starts to dominate at energies above 1~GeV. Since the latter configuration can
produce only two pions in its 
decay, the only resonance process eligible for four-pion production is the
double $N^*(1440)$ excitation, {\it i.e.} the $N^*(1440)N^*(1440)$ excitation
by $t$-channel meson exchange between the 
colliding incident nucleons. Also, the  nominal mass of $2m_{N^*(1440)}$ for this
configuration fits very well to the center-of-mass energies of the
measurements discussed here. A model calculation based on an extended
version of the modified Valencia model \cite{TS,Luis} reproduces the
the measured total cross sections within 30$\%$.

The solid lines in Fig.~3 show a calculation of the $N^*(1440)N^*(1440)$ 
process 
adjusted in height to the data. For the lower energy, $\sqrt s$ =
2.72 GeV, this calculation gives already a practically perfect description of
all three invariant-mass spectra within uncertainties.

\begin{figure} 
\centering
\includegraphics[width=0.99\columnwidth]{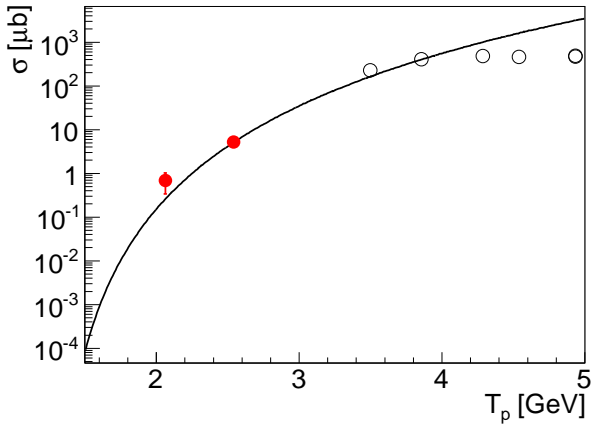}
\caption{\small (Color online) Energy dependence of the total cross section
  for the $pp \to pp\pi^+\pi^+\pi^-\pi^-$ reaction. The filled circles are
  from this work, the open symbols from
  Refs. \cite{alexan,danieli,ward,blob,blobel}. The drawn line gives the energy
  dependence of pure phase space normalized to the data point at $T_p$ = 2.541
  GeV.} 
\label{fig2}
\end{figure}

\begin{figure} [t]
\centering
\includegraphics[width=0.49\columnwidth]{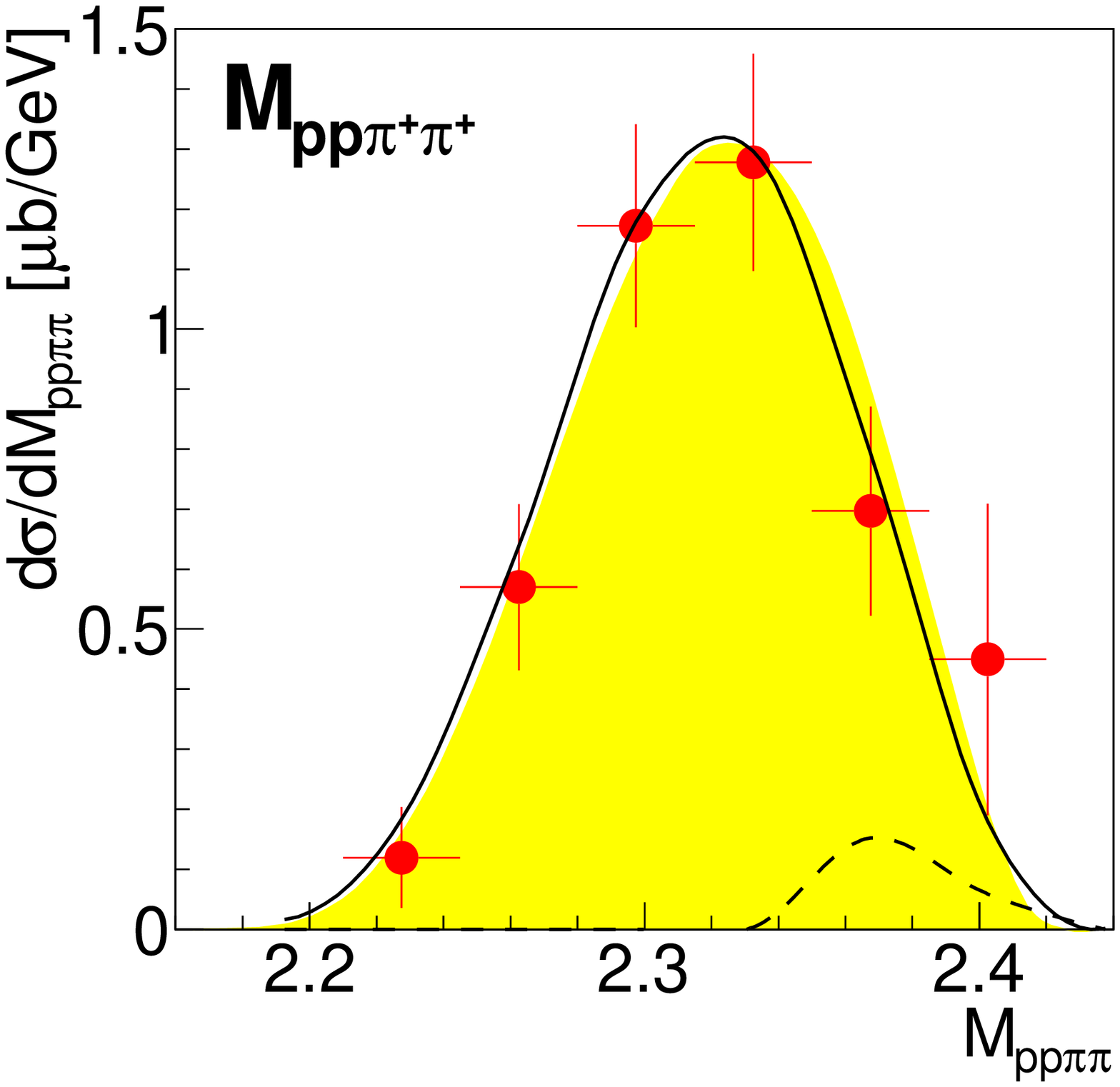}
\includegraphics[width=0.49\columnwidth]{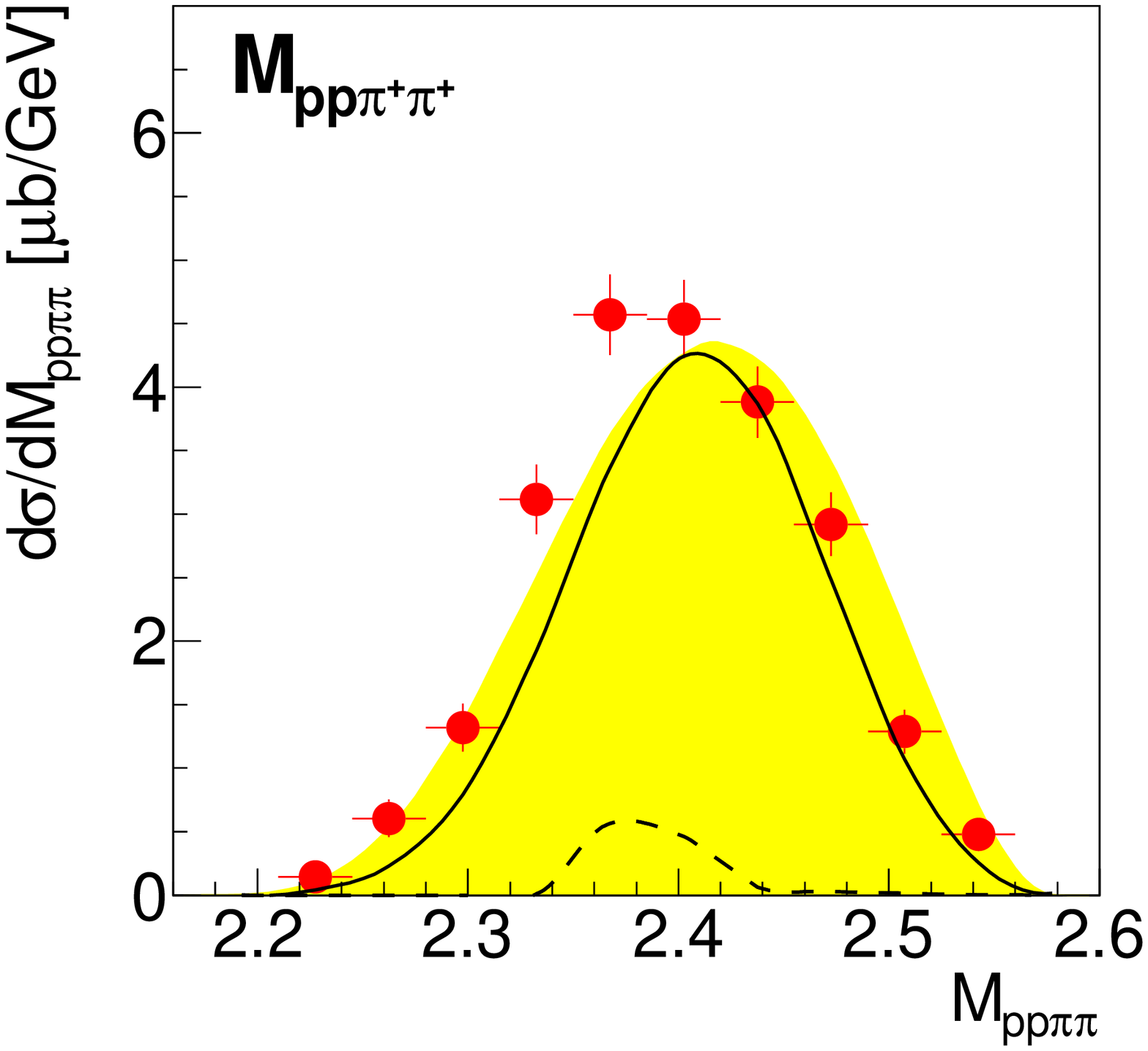}
\includegraphics[width=0.49\columnwidth]{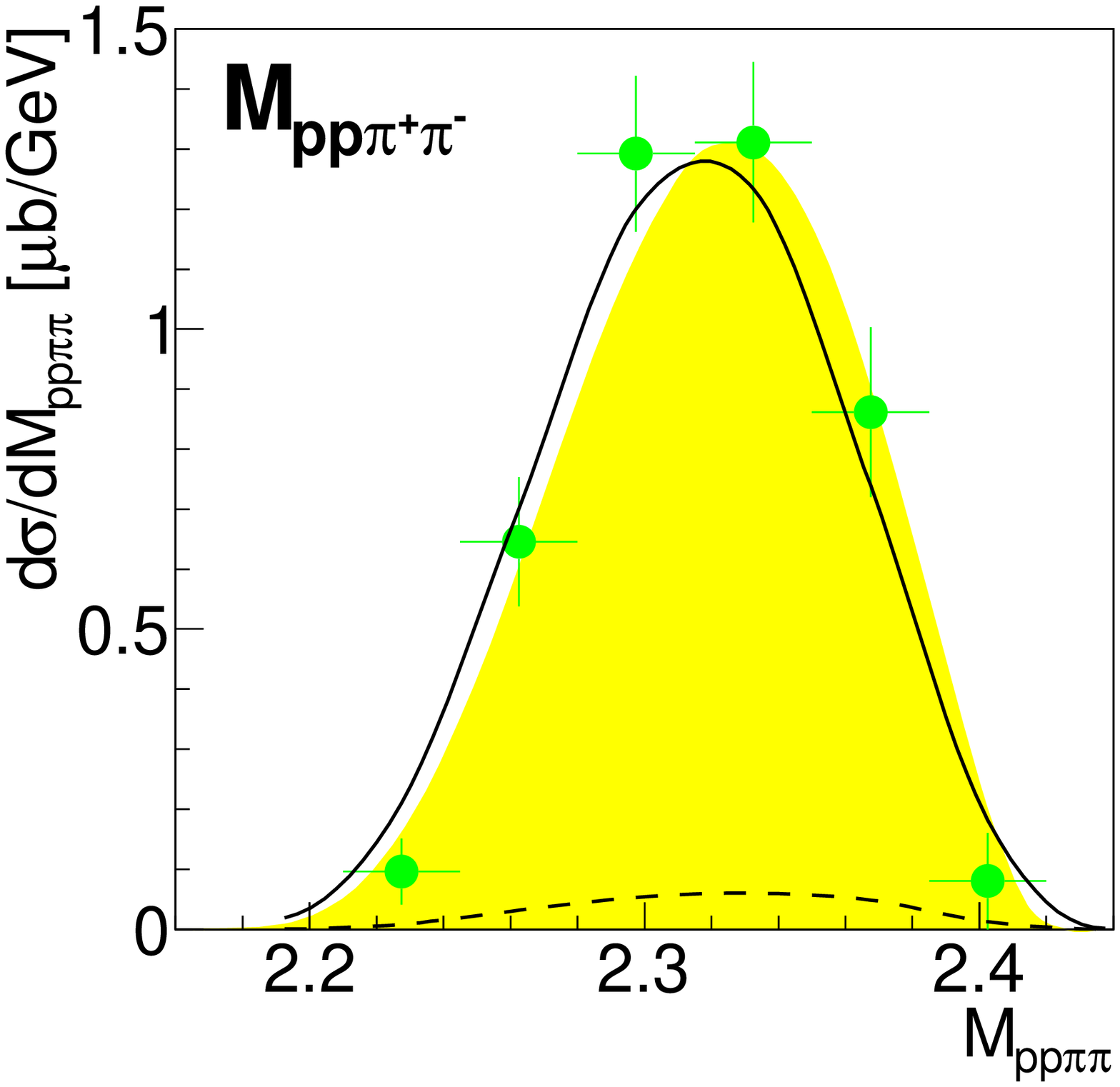}
\includegraphics[width=0.49\columnwidth]{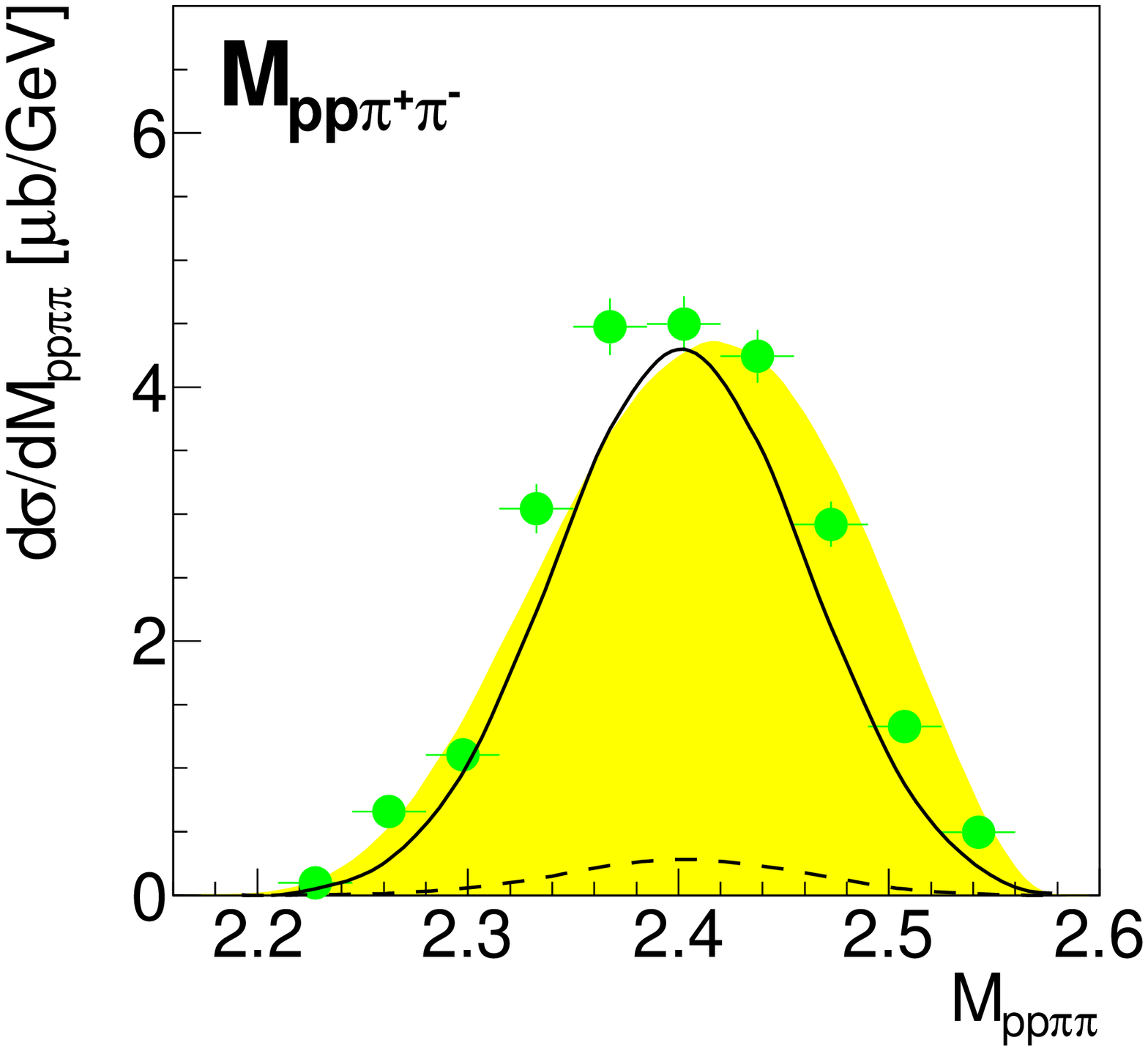}
\includegraphics[width=0.49\columnwidth]{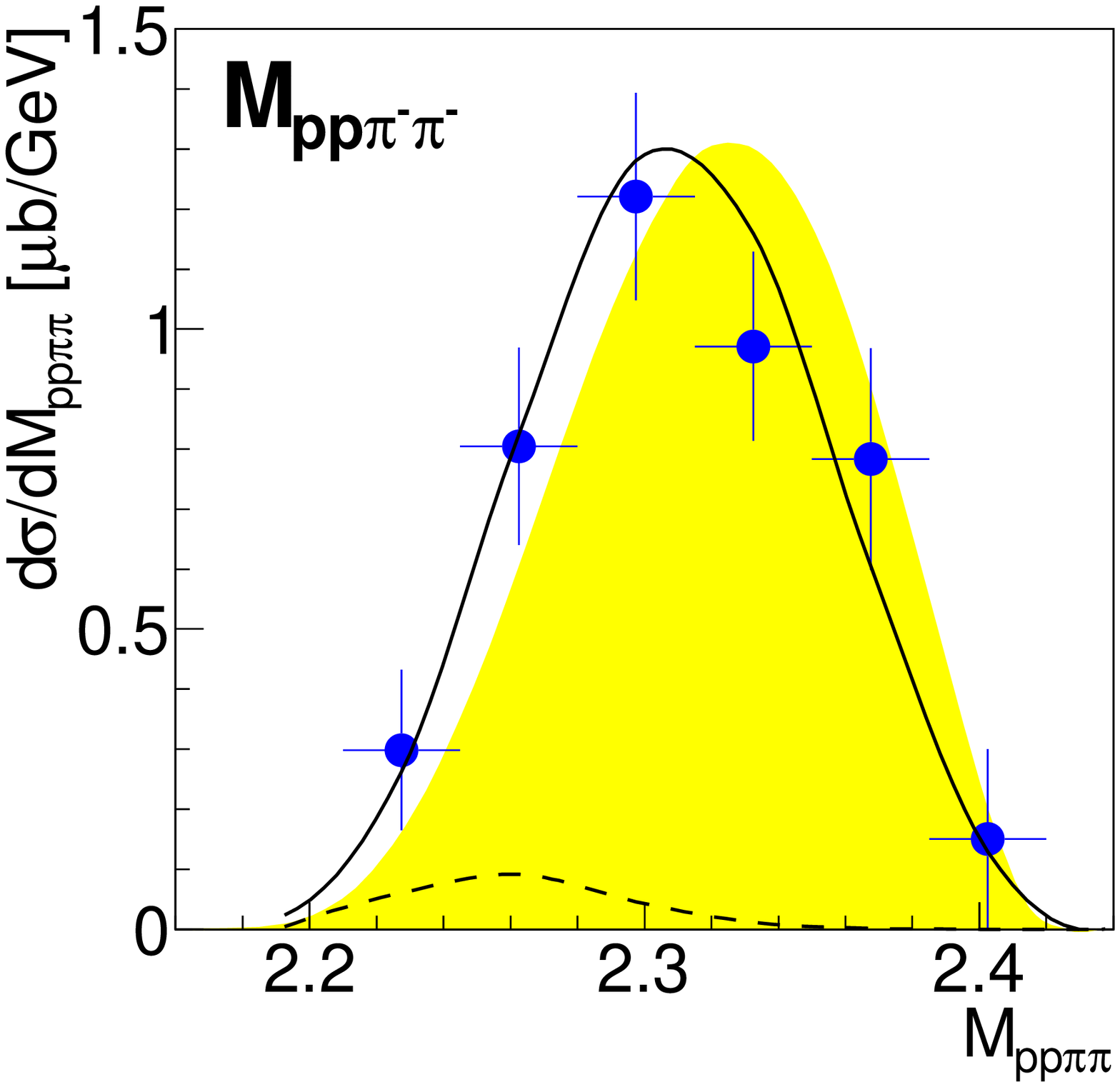}
\includegraphics[width=0.49\columnwidth]{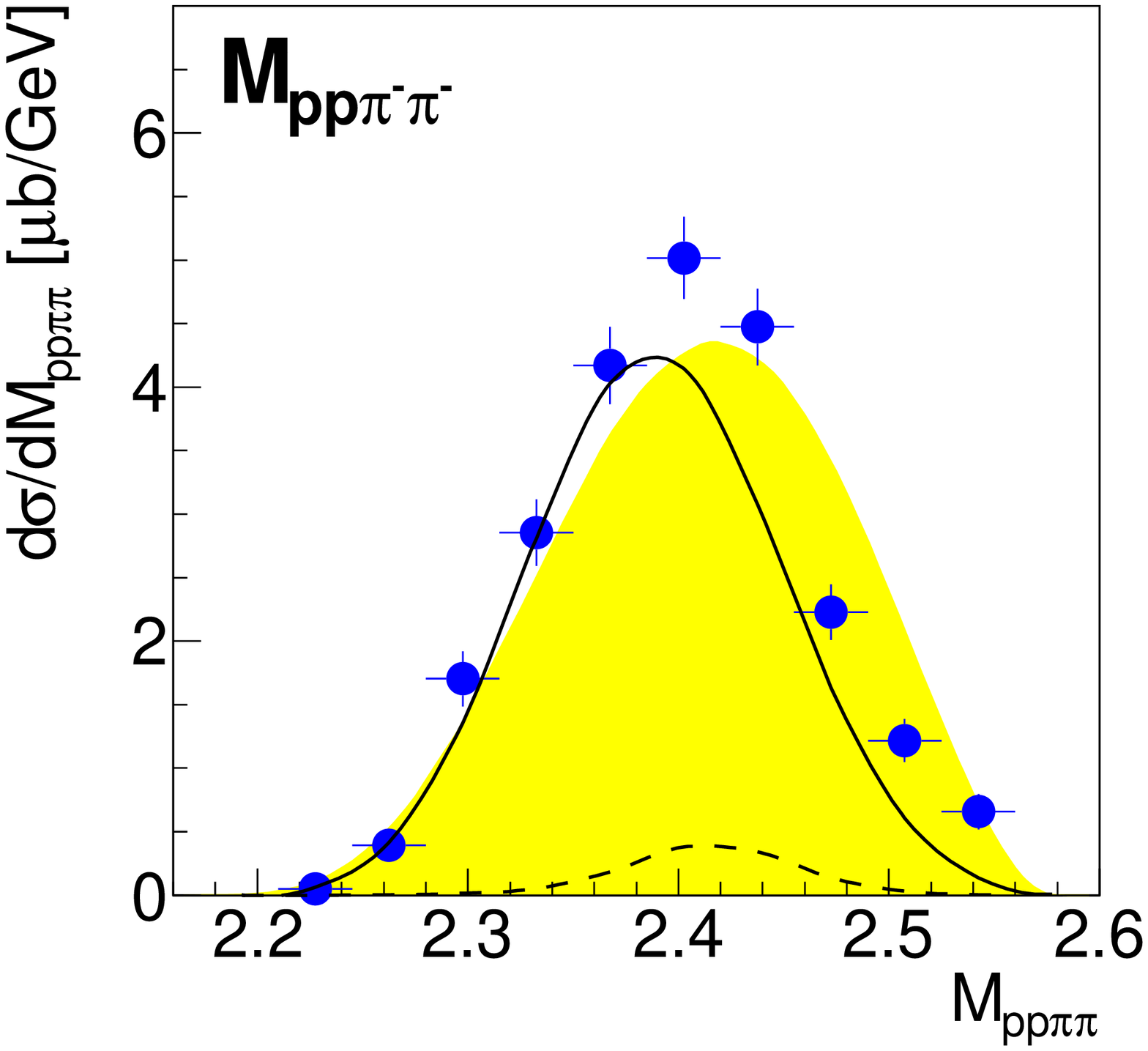}
\caption{\small (Color online) Distributions of invariant masses
  $M_{pp\pi^+\pi^+}$ (top), $M_{pp\pi^+\pi^-}$ (middle) and $M_{pp\pi^-\pi^-}$
  (bottom) for $\sqrt s$ = 2.72 GeV (left) and 2.88 GeV (right) within WASA
  acceptance. Solid dots
  denote the data from this work, the shaded histograms represent phase-space
  distributions, whereas the calculated $t$-channel $N^*(1440) N^*(1440)$
  distribution  is
  shown by the solid lines. 
  The dotted curves show the effect of an $I =3$
  resonance with mass m = 2380 MeV and width $\Gamma$ = 70 MeV scaled
  arbitrarily in height to a 5$\%$ contribution of the total cross section.
}
\label{fig3}
\end{figure}

\begin{figure}[t] 
\centering
\includegraphics[width=0.79\columnwidth]{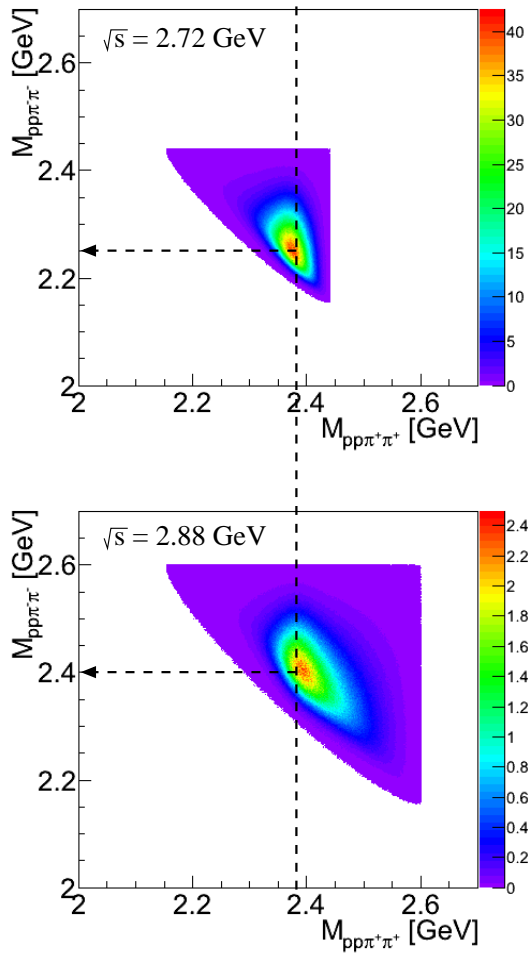}
\caption{\small (Color online) Distribution of MC-simulated events plotted in the
  plane of $M_{pp\pi^-\pi^-}$ versus $M_{pp\pi^+\pi^+}$ for an $I =3$
  resonance with mass m = 2380 MeV and width $\Gamma$ = 70 MeV. The top panel
  exhibits the situation at $\sqrt s$~=~2.72 GeV, the
  bottom panel that at $\sqrt s$~=~2.88 GeV.
}
\label{fig4}
\end{figure}

For the higher energy this description is not quite as good, since it misses
strength at low $pp\pi^+\pi^+$ and high $pp\pi^-\pi^-$ invariant masses. The
situation could possibly be improved, if we would fit a 
contribution from the next higher-lying $N^*$ excitation (providing a
$N^*(1520)N^*(1440)$ configuration in the intermediate state) to the data. But
we refrain here from a fine tuning of the background description due to the
the much increased complexity of the theoretical description, which
necessarily introduces new uncertainties. We just note 
that the behavior of the shapes given by the solid lines is characteristic for
a dominance of the $\Delta^{++}\Delta^{++}$ excitation insofar as the
$M_{pp\pi^+\pi^+}$ spectrum is narrower than the pure phase-space spectrum and
peaking around $2m_\Delta$, whereas the $M_{pp\pi^-\pi^-}$ spectrum
exhibiting dominantly the reflection of the $\Delta^{++}\Delta^{++}$
excitation peaks at substantial lower mass.

Next, we investigate, how a $I =3$ resonance would show up in these
spectra. From isospin coupling arguments, we deduce the relative cross sections,
with which such a resonance should show up in the various invariant-mass
spectra, namely: 
\begin{equation}
\sigma_{pp\pi^+\pi^+} : \sigma_{pp\pi^+\pi^-} :
\sigma_{pp\pi^-\pi^-} = 1 : \frac 2 {225} : \frac 1 {225}. 
\end{equation}

{\it I.e.}, such a resonance contributes practically only to the spectrum
with the highest charge in a direct way. However, since the three
invariant-mass spectra are interrelated, reflections of such a resonance also
appear in the $M_{pp\pi^+\pi^-}$ and $M_{pp\pi^-\pi^-}$ spectra -- as
illustrated in Fig.~4, where MC generated events are plotted in
the plane $M_{pp\pi^-\pi^-}$ versus $M_{pp\pi^+\pi^+}$. It displays the
results of a 
simulation of an $I = 3$ resonance with $m$ = 2380 MeV and $\Gamma$~=~70
MeV. The simulated resonance is shown in Fig.~3 by the dashed curves scaled
in height corresponding to a 5$\%$ contribution of the resonance to the total
cross section.  
Whereas in the
$M_{pp\pi^+\pi^-}$ spectrum the reflection causes a broad phase-space like
continuum, it produces a peak-like structure in the $M_{pp\pi^-\pi^-}$
distribution, though somewhat broader than the original peak in the
$M_{pp\pi^+\pi^+}$ spectrum and located in the complementary region of the
kinematical mass range. 

Knowing now the kinematic behavior of such an $I~=~3$ resonance, we further
inspect the data shown in Fig.~3. We observe no obvious narrow structures,
which fulfill the kinematical conditions for a possible $I = 3$ resonance.
However, we immediately also notice that a contribution of a dibaryon
resonance as 
illustrated by the dashed lines in Fig. 3 would certainly give an improved
description of the data. Though this is certainly a 
model-dependent statement, it demonstrates the difficulty of excluding a
dibaryon resonance contribution of smaller than 5$\%$ of the total cross
section -- in particular for dibaryon masses smaller than 2380 MeV.

In an ideal case the peak to be searched for is expected to sit upon a 
   flat or smoothly rising or falling background with a curvature, which is
   small compared to the peak width. This is far from being the case here. On
   theoretical grounds we can not expect the dibaryon resonance to have a
   width much smaller than 50 MeV, more likely is a  width in the region of
   100 MeV or even above, if this resonance happens to be  close to the
   $\Delta\Delta$ threshold. The background due to conventional processes is
   not flat or smoothly rising / falling in the range of interest as we see
   from the distributions displayed in  Fig.~3. Moreover shape and strength 
   of the background can not be calculated sufficiently reliable within
   contemporary theoretical approaches. Though the width of these distributions
   is still broader than the dibaryon signal we look for, it is not broader by
   an order of  magnitude. We are not aware of any model-independent peak
   search analysis for  such a case. Hence we will proceed by assuming two
   scenarios, where the background is accounted for either by phase space-like
   processes (meaning processes, which give identical contributions in all
   $M_{NN\pi\pi}$ spectra, e.g. chiral terms, various contact terms, etc...)
   or by the $N^*(1440)N^*(1440)$ process displayed in Fig.~3. The
   $N^*(1440)N^*(1440)$ scenario represents a theoretically-motivated
   background description, though possibly oversimplified as discussed above. 

As usual in such peak searches, we assume interferences to be small and add
   the resonance term incoherently to the background term. Under these
   assumptions the shapes of both resonance and background can be considered
   to be known (also within WASA acceptance). Then for given mass and width of
   the resonance only the  relative contributions of resonance and background
   enter the simultaneous fit of all three invariant mass spectra. The upper
   limits (95$\%$ C.L.) resulting from these single-parameter fits are
   displayed in Fig.~5 in dependence of a hypothetical dibaryon mass
   $M_{dibaryon}$ for assumed resonance widths of 50 MeV (solid lines), 100
   MeV (dotted) and 150 MeV (dashed). 
   
   In order to investigate the case, where the background is assumed to be
    distributed phase-space like, we consider the following difference spectra
    constructed out of the three invariant-mass spectra:  
      $\sigma_{pp\pi^+\pi^+} - \sigma_{pp\pi^-\pi^-}$,
      $\sigma_{pp\pi^+\pi^+} - \sigma_{pp\pi^+\pi^-}$ and
      $\sigma_{pp\pi^-\pi^-} - \sigma_{pp\pi^+\pi^-}$,   
    since they have the advantage that there the contributions from
    phase-space like distributions cancel. Note that any possible 
    contaminations from misreconstructed background, like three-pion
    production with subsequent $\pi^0$ Dalitz decay, cancels out in the
    difference spectra as well.

\begin{figure} [t] 
\centering
\includegraphics[width=0.99\columnwidth]{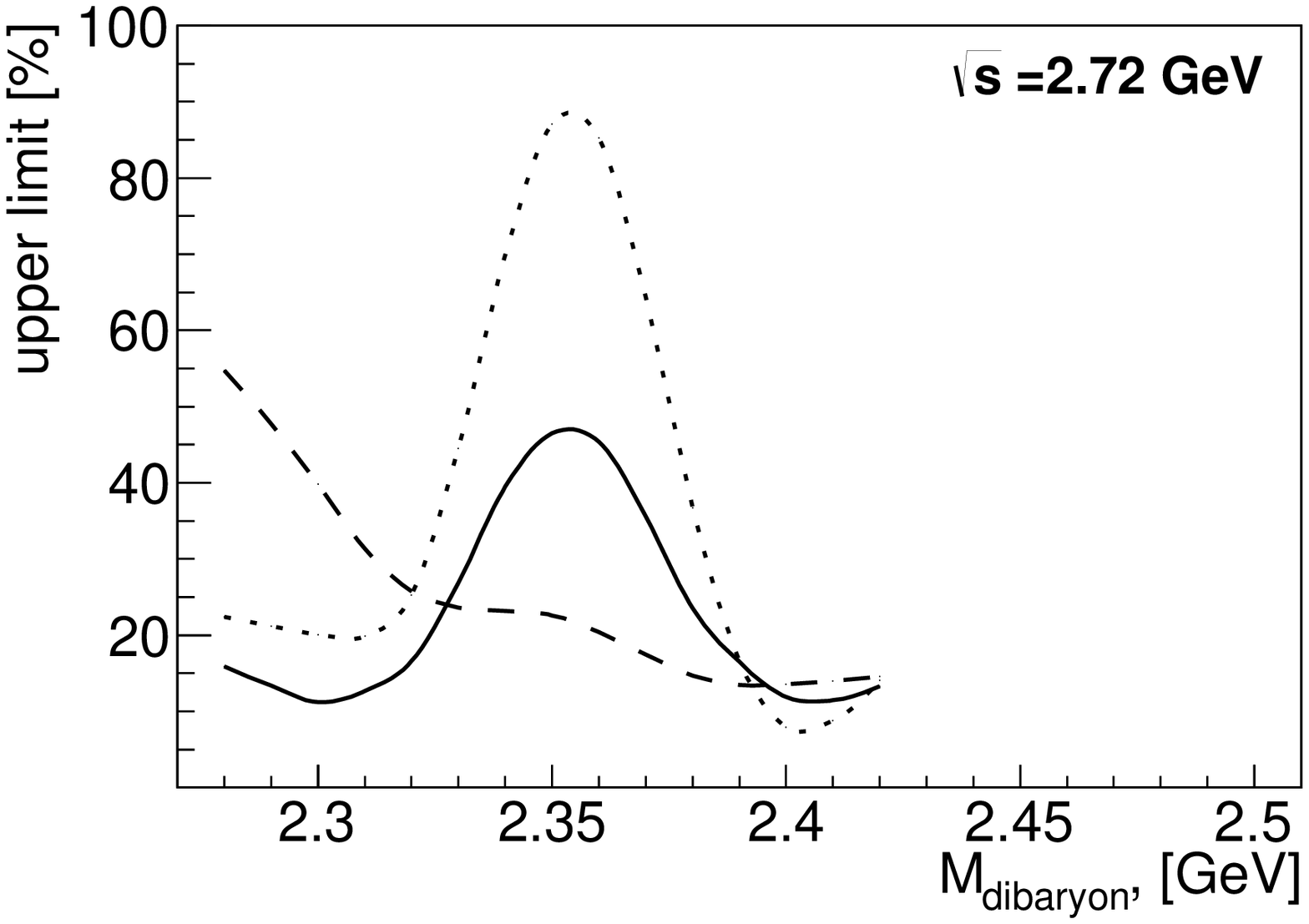}
\includegraphics[width=0.99\columnwidth]{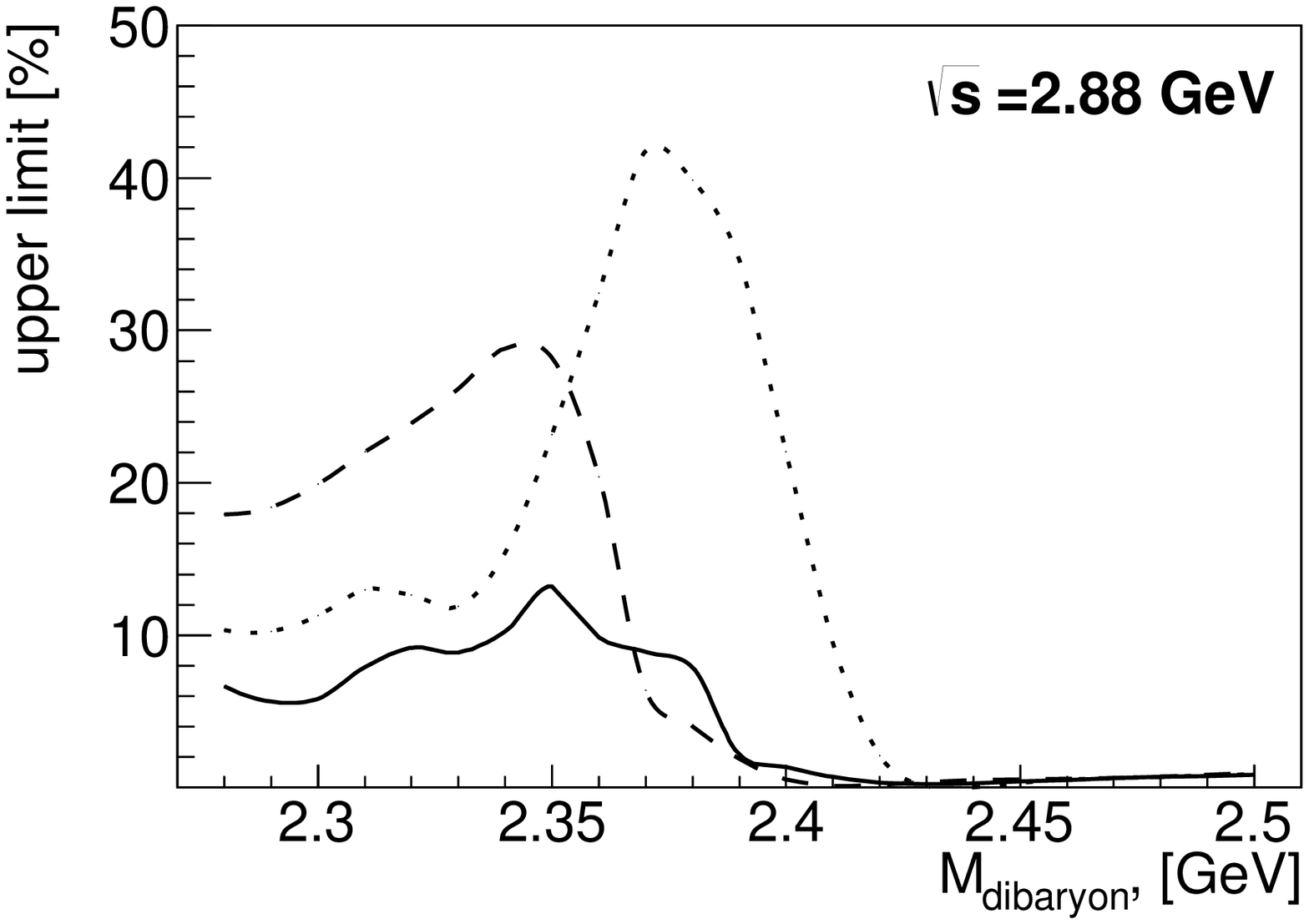}
\caption{\small (Color online) Upper limits (C.L. 95$\%$) in percentage of the total cross
  section from the search for a $I =3$
  resonance structure conducted on the invariant mass spectra spectra of
  Fig.~3 at $\sqrt s$ = 2.72 GeV (top) and 2.88 GeV (bottom) assuming the
  conventional processes to behave like the $N^*(1440)N^*(1440)$
  distributions. The  
  solid, dotted and dashed lines 
  refer to a fit search with a line width of $\Gamma$ = 50, 100 and 150 MeV,
  respectively. 
}
\label{fig6}
\end{figure}

In these difference spectra, which are plotted in Fig.~6 for both beam
energies, double baryon excitations due to
$t$-channel meson exchange produce 
an antisymmetric pattern (see solid lines in Fig.~6), whereas
an $I =3$ resonance in the $pp\pi^+\pi^+$ subsystem should show up in general by
an asymmetric pattern formed by its direct peak and its reflection
-- as indicated by the dashed curves in Fig.~6. 

\begin{figure}[t] 
\centering
\includegraphics[width=0.49\columnwidth]{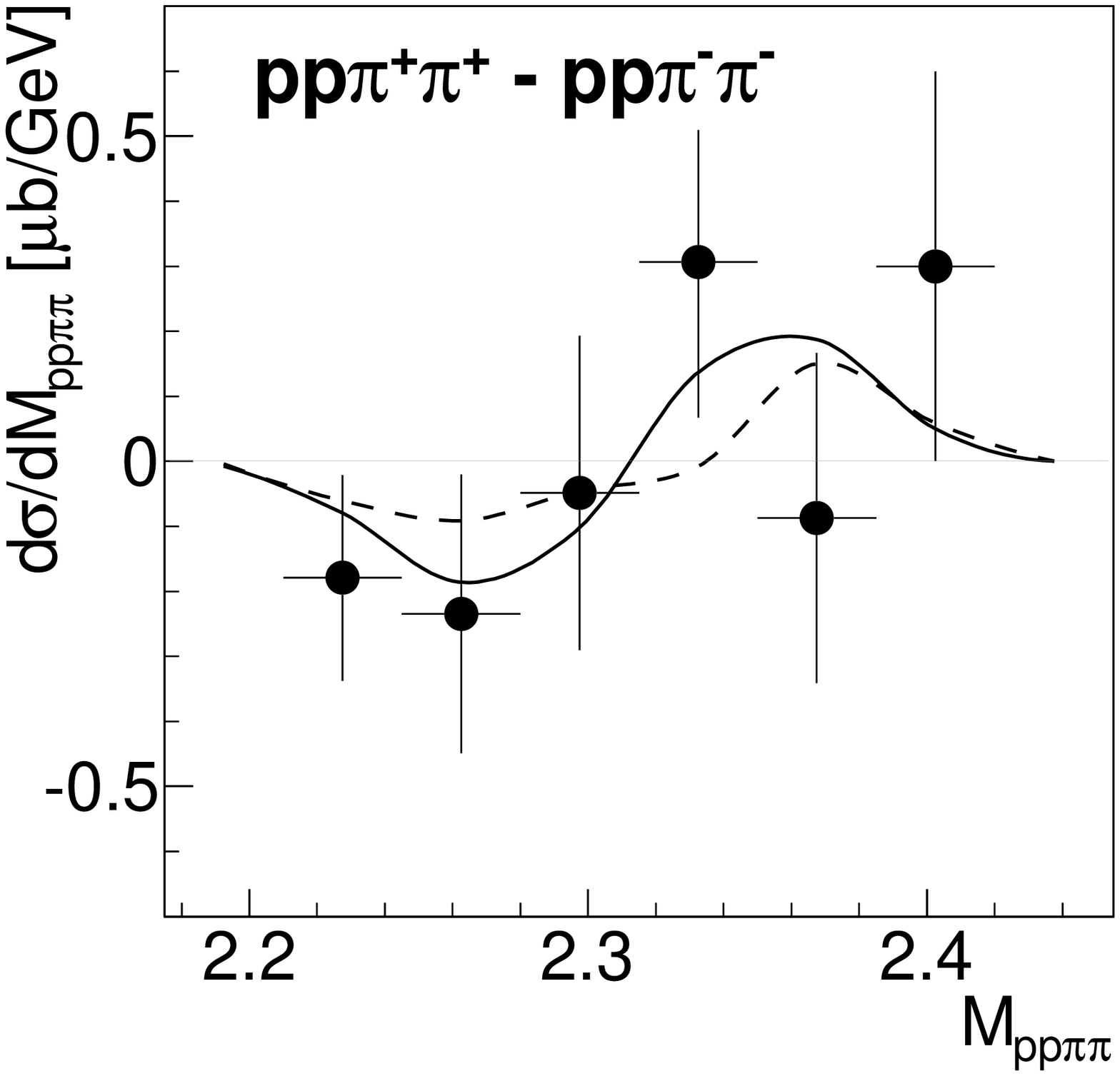}
\includegraphics[width=0.49\columnwidth]{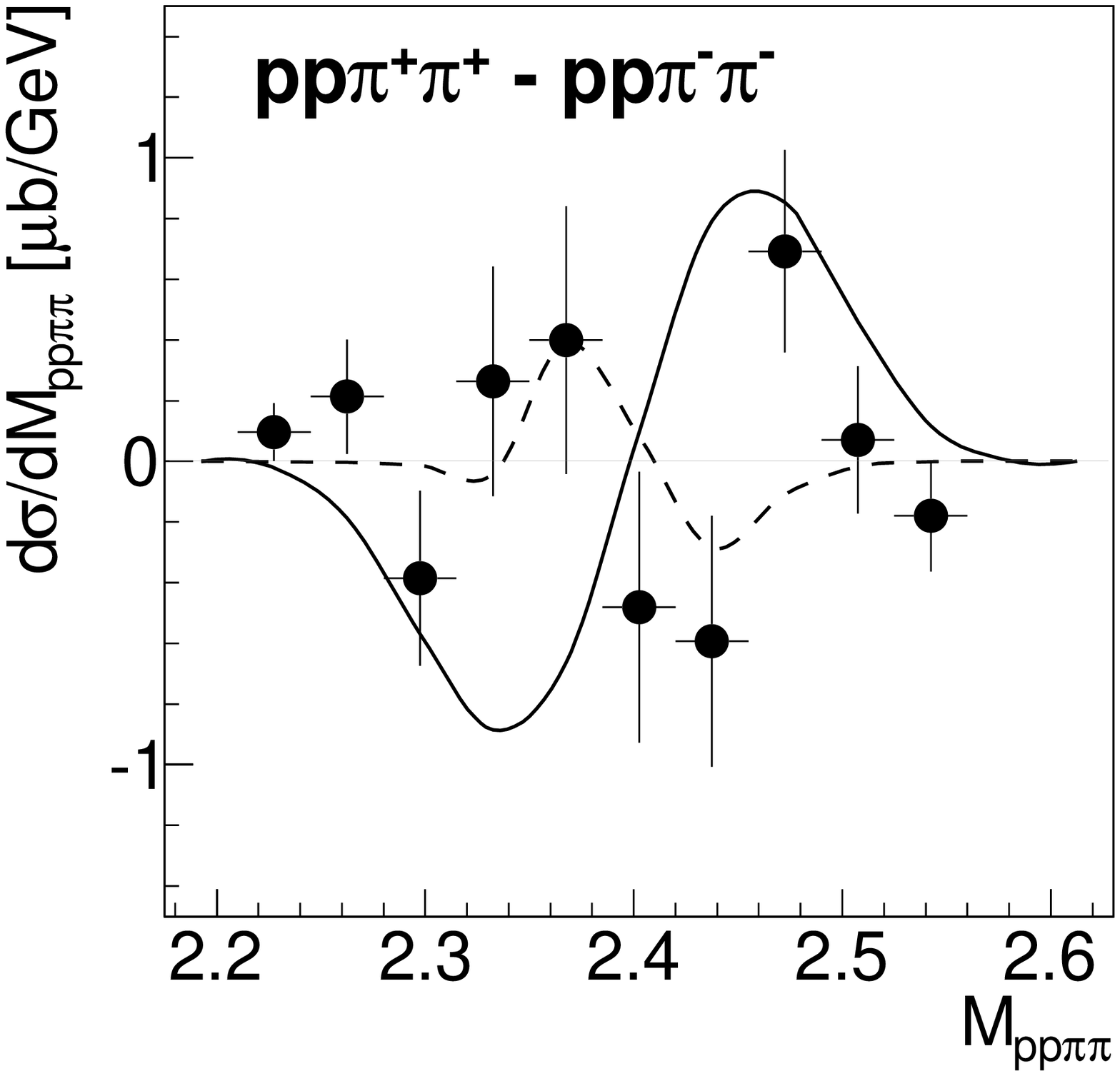}
\includegraphics[width=0.49\columnwidth]{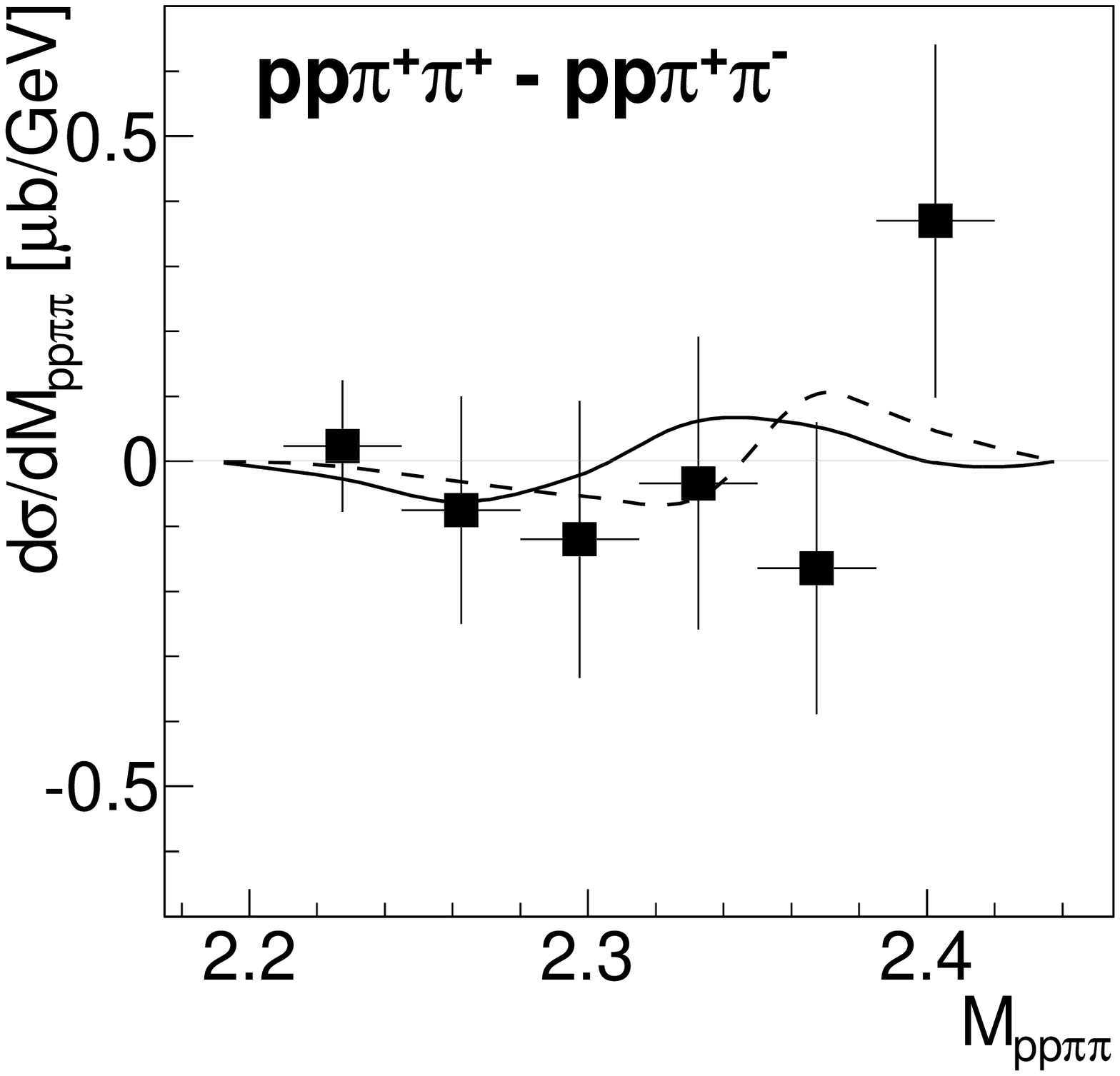}
\includegraphics[width=0.49\columnwidth]{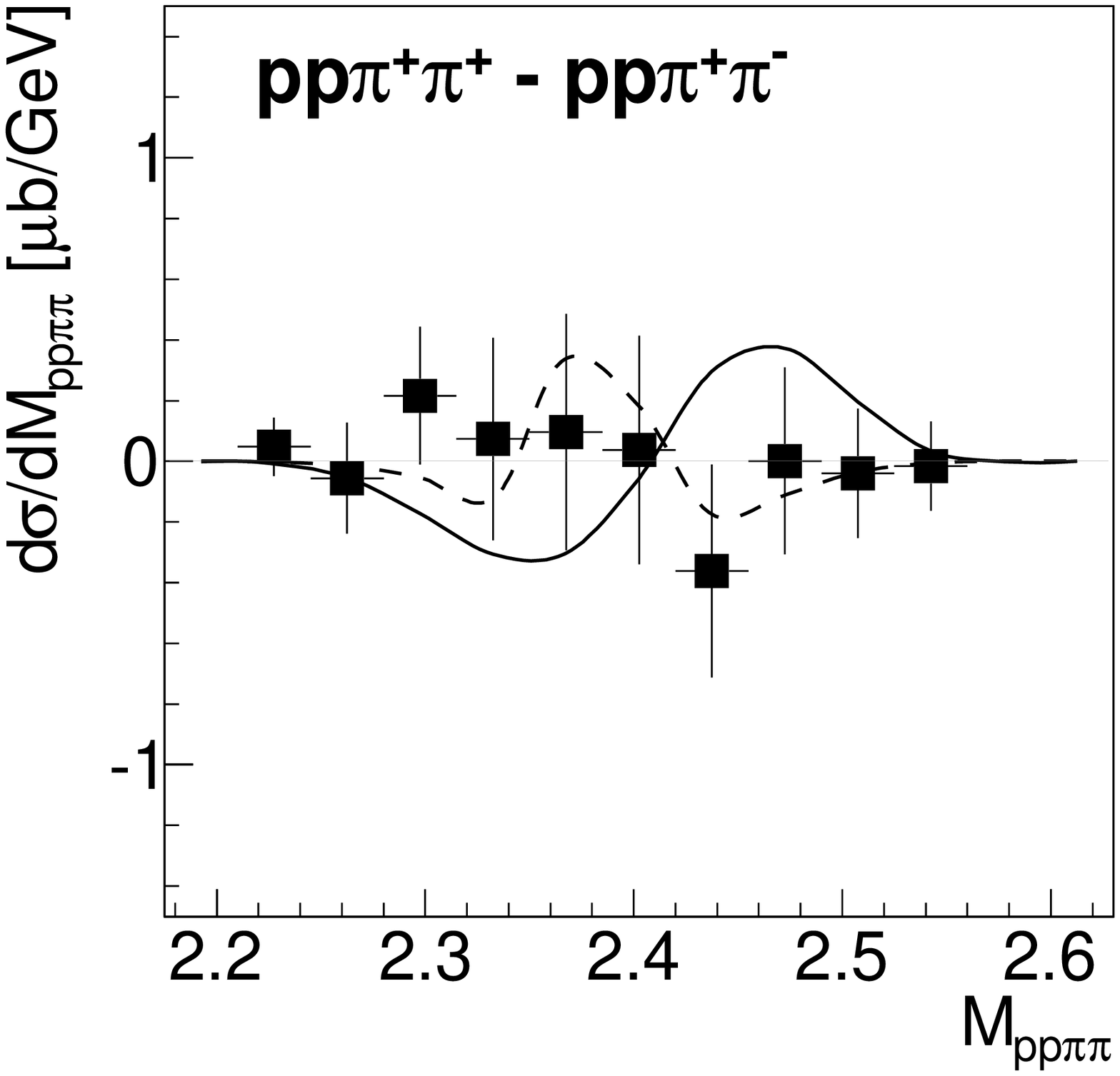}
\includegraphics[width=0.49\columnwidth]{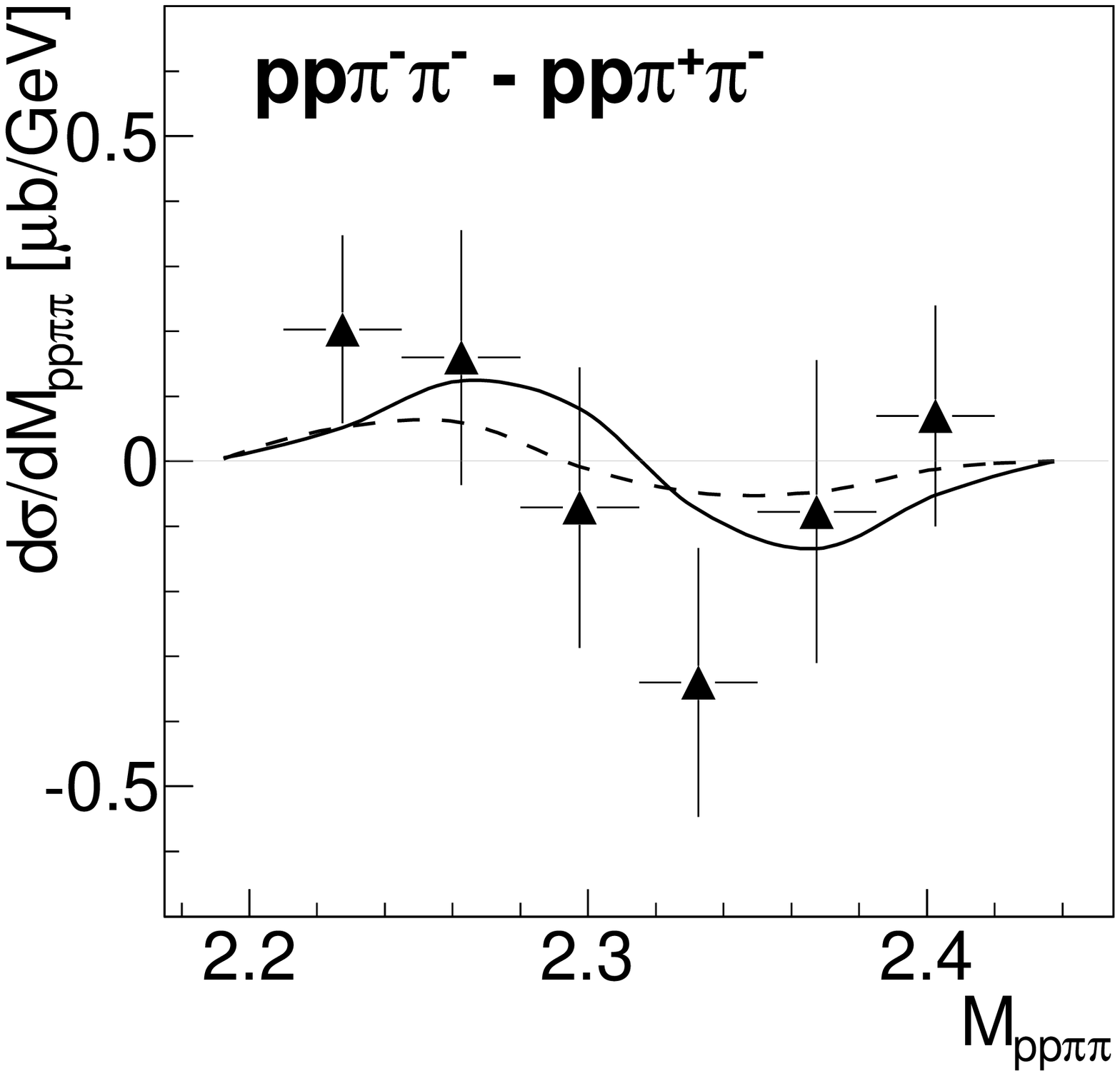}
\includegraphics[width=0.49\columnwidth]{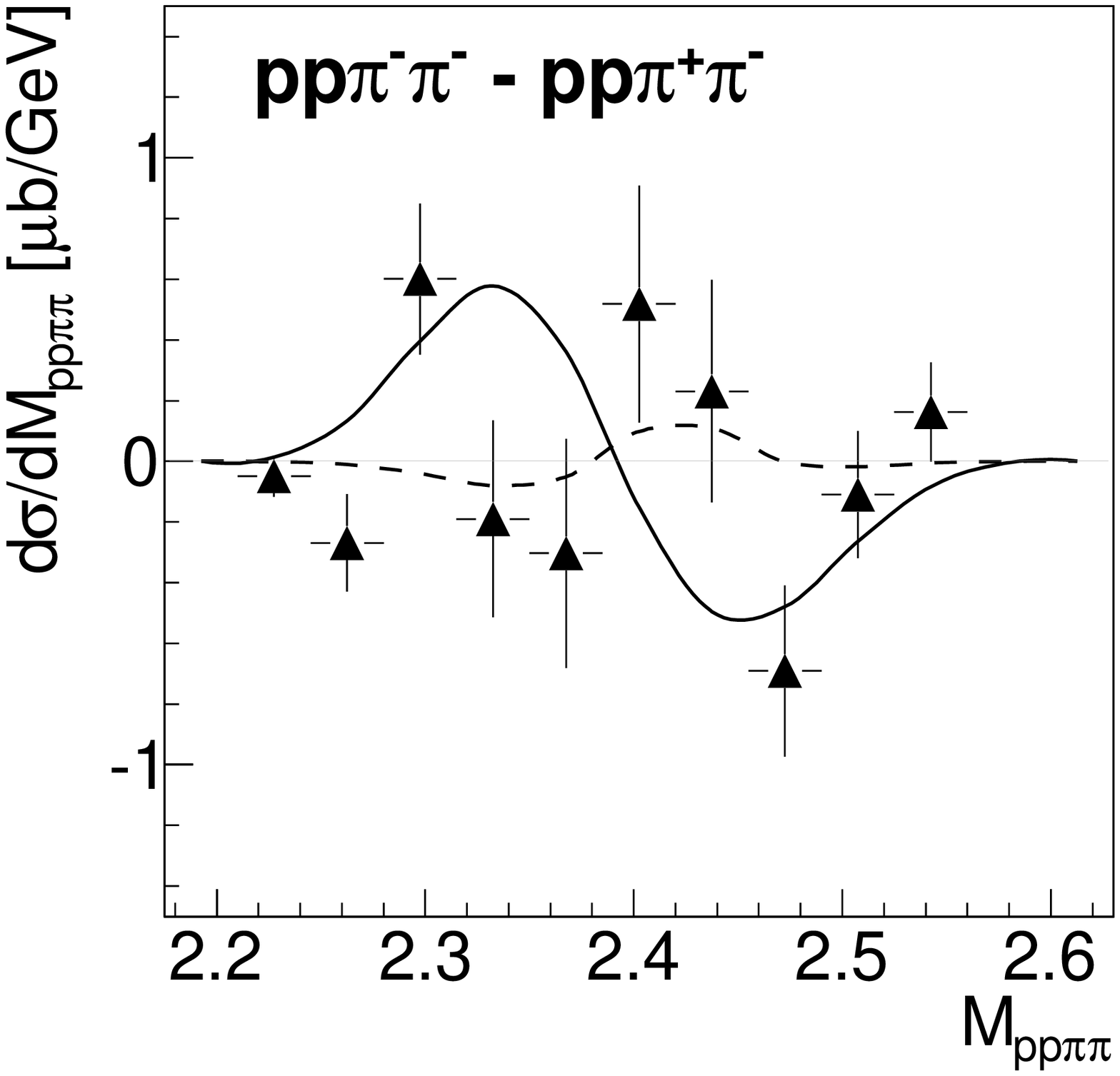}
\caption{\small (Color online) Difference spectra as defined in eqs. (2-4) in
  dependence of $Mpp\pi\pi$. The dashed curve represents the simulation of an
  $I =3$ resonance with mass $m$ = 2380 MeV and width $\Gamma$ = 70 MeV, the
  solid line  the $t$-channel $N^*(1440) N^*(1440)$ excitation. The left panel
  exhibits the situation at $\sqrt s$~=~2.72 GeV, the
  right panel that at $\sqrt s$~=~2.88 GeV.
}
\label{fig5}
\end{figure}

Since we know the expected signature of such a resonance in the difference
spectra, we can perform again single-parameter peak
finding fits simultaneously to all three difference spectra  per beam energy
and thus obtain upper limits for such a resonance in dependence of its mass
and width. The results for the 95$\%$ C.L. upper limits of this peak finding
search are displayed in Fig.~7. 




Both in Fig.~5 and in Fig.~7 the 95$\%$ C.L. upper limits
are plotted in percentage of the total cross section. 
The extrapolation of our results obtained within the WASA acceptance to total
cross sections introduces systematic uncertainties, as discussed in the
experimental section. They amount to 40$\%$ for the lower energy and 20$\%$
for the higher energy. 

For both scenarios -- $N^*(1440)N^*(1440)$ and phase-space like background --
we obtain qualitatively similar results. 
Due to the much superior statistics at the higher energy, the resulting upper
limits are much more stringent there. 
As expected, the data are most sensitive to the signature of a narrow
resonance. Also, for large dibaryon masses the upper limits are in general
substantially lower than for small masses. The largest
upper limit happens  in the $N^*(1440)N^*(1440)$ background scenario for a
dibaryon mass of about 2380 MeV and a width of 100 MeV, where the upper limit
reaches 40$\%$ of the total cross section.


Compared to the formation cross section of 1.7 mb found for
$d^*(2380)$ \cite{BR}, the upper limits found here for the production of an
$I=3$ dibaryon resonance are smaller by three to four orders of magnitude.

\begin{figure} [t] 
\centering
\includegraphics[width=0.99\columnwidth]{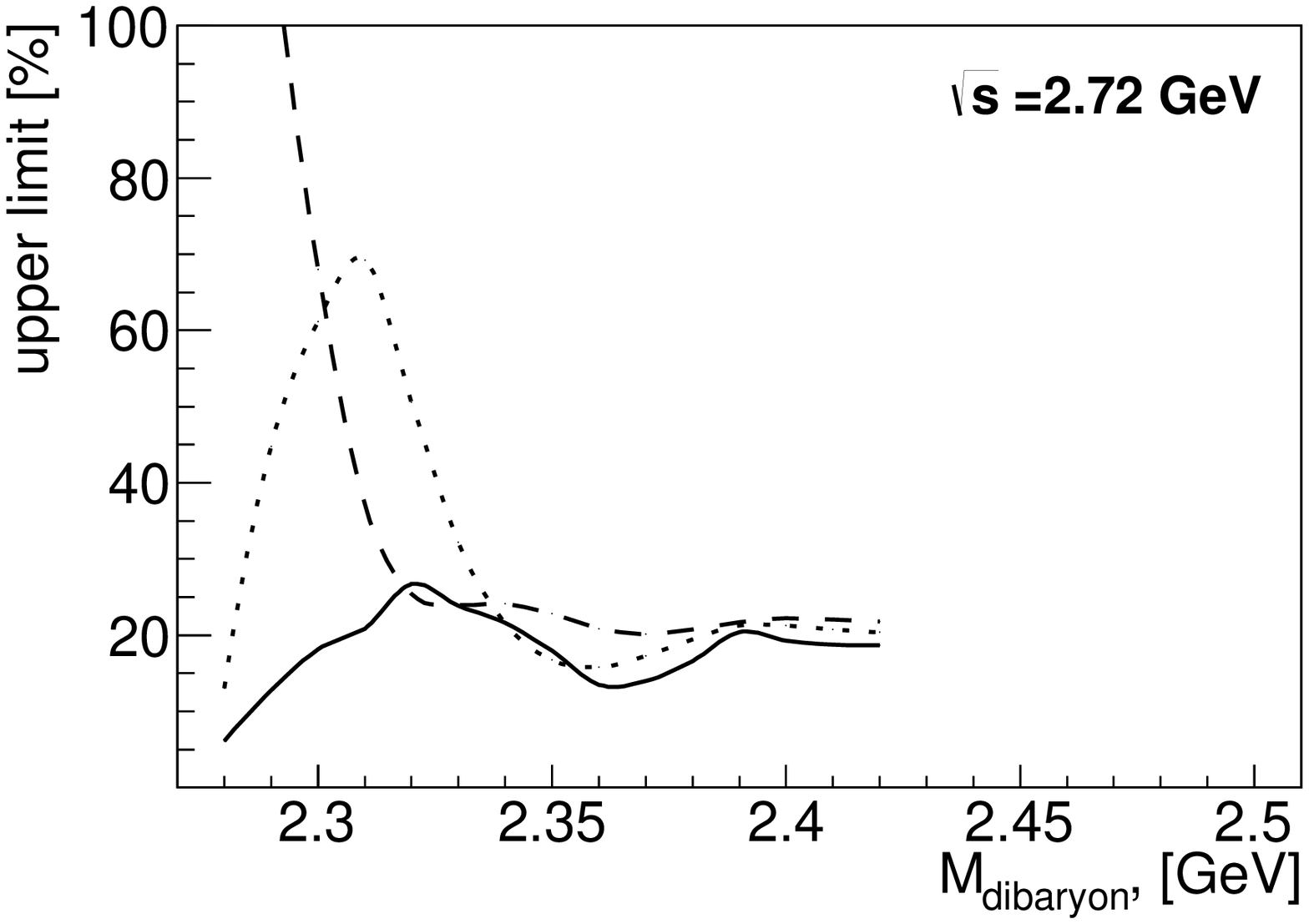}
\includegraphics[width=0.99\columnwidth]{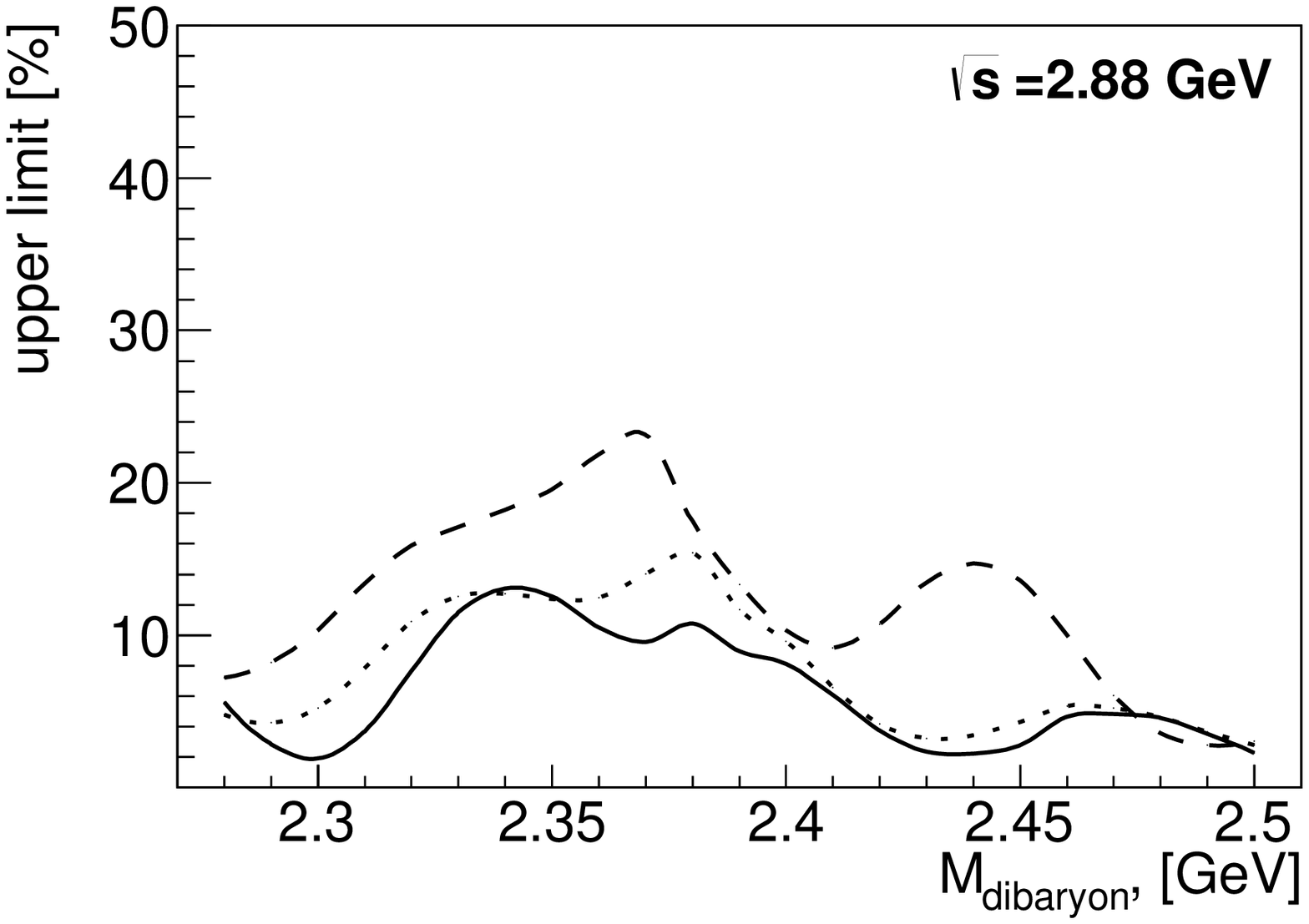}
\caption{\small (Color online) Upper limits (C.L. 95$\%$) in percentage of the total cross
  section from the search for a $I =3$
  resonance structure conducted on the difference spectra
  defined in eqs. (2-4) at $\sqrt s$ = 2.72 GeV (top) and 2.88 GeV (bottom). The
  solid, dotted and dashed lines 
  refer to a fit search with a peak width of $\Gamma$ = 50, 100 and 150 MeV,
  respectively. 
}
\label{fig6}
\end{figure}


More informative should be the comparison to formation / production of a
$\Delta\Delta$ system by conventional $t$-channel meson exchange. In 
two-pion production (isoscalar part) the peak cross section for $d^*(2380)$
formation is roughly one order of magnitude larger than the one for the
conventional $\Delta\Delta$ process \cite{MB} at the $d^*(2380)$ peak
energy. If we assume that the four-pion 
production at the beam energies considered here is dominated by
$N^*(1440)N^*(1440)$ formation as shown in Fig.~3, then we know also the cross
section for conventional $\Delta^{++}\Delta^{++}$ production via the route $pp
\to N^*(1440)N^*(1440) \to \Delta^{++}\Delta^{++}\pi^-\pi^-$. The
calculations shown in Fig.~3 contain the two-pion decay routes of the Roper
resonance $N^*(1440) \to N\sigma \to N\pi\pi$ and $N^*(1440) \to \Delta\pi \to
N\pi\pi$ with the branching ratio obtained in Refs.~\cite{TS,AA,PDG}. From
these calculations we find that at $\sqrt s$ = 2.72 GeV about 6$\%$ of the total
cross section 
 are due to conventional
$\Delta^{++}\Delta^{++}$ production. For $\sqrt s$~=~2.88 GeV the corresponding
number is 32$\%$. 

Whereas for the lower incident energy the upper limits obtained for $D_{30}$
production are in general larger than the cross section for conventional
$\Delta^{++}\Delta^{++}$ production, the
upper limits obtained at the higher incident energy are in general
significantly smaller -- with the exception of the case $M_{dibaryon} \approx$
2380 MeV and $\Gamma$ = 100 MeV for the $N^*(1440)N^*(1440)$ scenario, where
the upper limit is of the same order as the conventional
$\Delta^{++}\Delta^{++}$ production. The results for the higher incident
energy  appear to be quite significant. If the 
interaction between the two $\Delta^{++}$ particles produced side-by-side in
the decay of the intermediate $N^*(1440)N^*(1440)$ system would be attractive,
then the probability to form a dibaryon should be substantially larger than
for the conventional process  -- as it is obviously the case for
$d^*(2380)$ formation in the presence of an isoscalar $\Delta^+\Delta^0$
system. However, our results suggest that the probability for dibaryon
formation in the presence of a $\Delta^{++}\Delta^{++}$ system in the
intermediate state is smaller (with the possible exception of the above
mentioned case). This is in support
of the findings of Ref.~\cite{goldman}, which predicted an attractive
interaction between the $\Delta\Delta$ pair in case of $d^*(2380)$, but
repulsion in case of $D_{30}$ and hence no dibaryon formation.

\section{Summary and Conclusions}

We have searched for a $I = 3$ dibaryon resonance, which has been predicted by
Dyson and Xuong as well as by various QCD-based and hadronic model
calculations to decay 
into the $NN\pi\pi$ system. The mass range of our search covers the region
from 2.2 - 2.5 GeV, {\it i.e.} from near-to two-pion threshold to the nominal
$\Delta\Delta$ threshold of $2m_\Delta$ and above. To our knowledge this has
been the first such search -- with the exception of some earlier attempt by
use of proton-nucleus collisions \cite{Igor}.

We have found no apparent
indication for such a resonance in our data. The deduced upper cross section
limits for 
the production of such a resonance are three to four orders of magnitude
smaller than the formation cross section of 1.7 mb found for $d^*(2380)$. They
also are up to one order of magnitude smaller than the cross section for
conventional $\Delta^{++}\Delta^{++}$ production in the $pp \to
pp\pi^+\pi^+\pi^-\pi^-$ reaction -- again in sharp contrast to
the corresponding situation for $d^*(2380)$ formation, where this is an order of
magnitude larger than in conventional $\Delta\Delta$ formation. 

An improved, reliable background description by conventional $t$-channel meson
exchange processes would certainly have the potential to lower these upper
limits considerably.

With only upper limits at present we, of course,  cannot exclude the existence
of such a 
resonance. However, if existent, either the production process of the $I = 3$
resonance associated with the emission of two pions has an unusually small
cross section or such a resonance has a mass above the energy region
investigated here -- as predicted, {\it e.g.} in Ref. \cite{mulders}. However,
in such a case, when the resonance lies significantly above  
the $\Delta\Delta$ threshold, its width is expected to be very broad due to its
fall-part decay and hence it will be very hard to distinguish such a resonance
from conventional processes.

\section{Acknowledgments}

We acknowledge valuable discussions with St. Brodsky,  A. Gal, I. Strakovsky,
F. Wang, C. Wilkin and
Z. Zhang on this issue. We are indebted to Luis Alvarez-Ruso for using his code.
This work has been supported by DFG (CL 214/3-1), STFC (ST/L00478X/1) and by
the Polish National Science Center through grants No. DEC-2013/11/N/ST2/04152
and 2011/03/B/ST2/01847.

\end{document}